\documentclass[twocolumn,aps,prb,superscriptaddress,floatfix]{revtex4-2}
\usepackage{amsmath}
\usepackage{amssymb}
\usepackage{graphicx}
\usepackage{color}

\newcommand{\bea}{\begin{eqnarray}}
\newcommand{\eea}{\end{eqnarray}}
\newcommand{\bse}{\begin{subequations}}
\newcommand{\ese}{\end{subequations}}

\newcommand{\tcs}{${\rm ThCr_2Si_2}$}
\newcommand{\cca}{CaCo$_{2-y}$As$_2$}
\newcommand{\sca}{${\rm SrCo_2As_2}$}

\newcommand{\bca}{${\rm BaCo_2As_2}$}

\newcommand{\cna}{${\rm CaNi_2As_2}$}
\newcommand{\csca}{Ca$_{1-x}$Sr$_x$Co$_{2-y}$As$_2$}
\newcommand{\scna}{Sr(Co$_{1-x}$Ni$_x$)$_2$As$_2$}
\newcommand{\ccia}{Ca(Co$_{1-x}$Ir$_x$)$_{2-y}$As$_2$}
\newcommand{\ccfa}{Ca(Co$_{1-x}$Fe$_x$)$_{2-y}$As$_2$}
\newcommand{\ccna}{Ca(Co$_{1-x}$Ni$_x$)$_{2-y}$As$_2$}
\newcommand{\bcia}{Ba(Co$_{1-x}$Ir$_x$)$_{2}$As$_2$}
\newcommand{\ecna}{Eu(Co$_{1-x}$Ni$_x$)$_{2-y}$As$_2$}
\begin{document}

\title{Suppression of antiferromagnetic order and strong ferromagnetic spin fluctuations in Ca(Co$_{1-x}$Ni$_x$)$_{2-y}$As$_2$ single crystals}

\author{Santanu Pakhira}
\affiliation{Ames Laboratory, Iowa State University, Ames, Iowa 50011, USA}
\author{Y. Lee}
\affiliation{Ames Laboratory, Iowa State University, Ames, Iowa 50011, USA}
\author{Liqin Ke}
\affiliation{Ames Laboratory, Iowa State University, Ames, Iowa 50011, USA}
\author{V. Smetana}
\affiliation{Department of Materials and Environmental Chemistry, Stockholm University, Svante Arrhenius v\"{a}g 16 C, 106 91 Stockholm, Sweden}
\author{A.-V. Mudring}
\affiliation{Department of Materials and Environmental Chemistry, Stockholm University, Svante Arrhenius v\"{a}g 16 C, 106 91 Stockholm, Sweden}
\author{Thomas Heitmann}
\affiliation{The Missouri Research Reactor and Department of Physics and Astronomy, University of Missouri, Columbia, Missouri 65211, USA}
\author{David Vaknin}
\affiliation{Ames Laboratory, Iowa State University, Ames, Iowa 50011, USA}
\affiliation{Department of Physics and Astronomy, Iowa State University, Ames, Iowa 50011, USA}
\author{D. C. Johnston}
\affiliation{Ames Laboratory, Iowa State University, Ames, Iowa 50011, USA}
\affiliation{Department of Physics and Astronomy, Iowa State University, Ames, Iowa 50011, USA}

\date{\today}

\begin{abstract}

\cca\ is a unique itinerant system having strong magnetic frustration. Here we report the effect of electron doping on the physical properties resulting from Ni substitutions for Co.  The single crystals of \ccna\ were characterized by single-crystal x-ray diffraction, energy-dispersive x-ray spectroscopy, magnetization $M$ versus temperature~$T$, magnetic field~$H$, and time~$t$, and heat capacity $C_{\rm p}(H,T)$ measurements. The A-type antiferromagnetic (AFM) transition temperature~$T_{\rm N}=52$~K for $x=0$ decreases to 22~K with only 3\% Ni substitution and is completely suppressed for $x > 0.11$.  For $0.11\leq x \leq0.52$ strong ferromagnetic~(FM) fluctuations develop as revealed by magnetic susceptibility $\chi(T) = M(T)/H$ measurements.  For $x = 0.11$ and 0.16, competing AFM and FM interactions result in spin-glass behavior at low~$T$ as evidenced by observations of thermomagnetic hysteresis and magnetic relaxation. Enhanced FM fluctuations are also found for the $x = 0.21$ and 0.31 crystals, where $\chi_c$ increases significantly at low~$T$\@. A large $\chi$ anisotropy in these compositions where $\chi_c$ is up to a factor of two larger than $\chi_{ab}$ suggests that the FM spin fluctuations are quasi-1D in nature. Weak FM contributions to $M(H=0)$ were found at $T=2$~K for $x=0.11$--0.31.  Heat-capacity $C_{\rm p}(T)$ measurements revealed the presence of FM quantum spin fluctuations for $0.11 \leq x \leq 0.52$, where a logarithmic $T$ dependence of $C_{\rm p}(T)/T$ is observed at low~$T$\@. The suppression of AFM order by the development of strong FM fluctuations in \ccna\ crystals suggests the presence of a FM quantum-critical point  at $x \approx 0.20$.  Our density-functional theory (DFT) calculations confirm that FM fluctuations are enhanced by Ni substitutions for Co in \cca.  The Sommerfeld electronic heat-capacity coefficient is enhanced for $x =0$, 0.21, and 0.42 by about a factor of two compared to DFT calculations of the bare density of states (DOS) at the Fermi energy, suggesting an enhancement of the DOS from electron-phonon and/or electron-electron interactions. The crystals with $x > 0.52$ do not exhibit FM  spin fluctuations or magnetic order at $T\geq 1.8$~K, which was found from the DFT calculations to result from a Stoner transition.  Superconductivity is not observed above 1.8~K for any of the compositions.  Neutron-diffraction studies of crystals with $x=0.11$ and~0.16 in the crossover regime $0.1\lesssim x \lesssim 0.2$ found no evidence of A-type ordering above 4.8~K within experimental resolution as observed in the parent compound with $x=0$.  Furthermore, no other common magnetic structures, such as  FM, helical stacking of in-plane FM layers, or in-plane AFM structure, were found above 4.8~K with an ordered moment greater than the uncertainty of 0.05~$\mu_{\rm B}$ per transition-metal atom.

\end{abstract}

\maketitle

\section{Introduction}

The discovery of superconductivity in doped iron arsenides by suppressing the long-range antiferromagnetic (AFM) order of parent compounds has opened up a wide research field studying the complex interplay between superconductivity (SC) and magnetism~\cite{Rotter2008, Sasmal2008, Jasper2008, Kumar2009, Ren2009}. A considerable number of studies have been carried out to understand the mechanism of the observed high-$T_c$ superconductivity in doped \tcs-structure  (122-type) $A$Fe$_2$As$_2$ ($A$ = Ca, Sr, Ba, Eu) compounds~\cite{Johnston2010, Paglione2010, Canfield2010, Stewart2011, Dai2012, Dai2015}. It has been established that AFM fluctuations mediate the electron-pairing mechanism for SC in the iron-arsenide family of compounds~\cite{Johnston2010, Dai2015, Scalapino2012}. Depending on the ratio of tetragonal lattice parameters $c/a$, a collapsed-tetragonal (cT) or uncollapsed-tetragonal (ucT) version of the structure is formed~(for a review, see Ref.~\cite{Anand2012}).

In contrast to 122-type iron-arsenide compounds, none of the parent cobalt arsenides exhibit superconductivity even with doping but are still found to be rich in physics associated with itinerant magnetism of the conduction electrons~\cite{Quirinale2013, Anand2014Ca, Pandey2013, Bing2019sca, Jayasekara2013, Wiecki2015, Li2019, Sefat2009, Anand2014, Xu2013, Sangeetha_EuCo2As2_2018}.  Interesting physical properties are observed including non-Fermi-liquid behavior associated with a quantum critical point, spin-glass behavior, and helical magnetic ordering in electron- or hole-doped 122-type CoAs-materials~\cite{Sangeetha2017, Jayasekara2017, Sangeetha2019scna, YLi2019, Wilde2019, Santanu2020CCIA, Santanu2021BCIA, Sangeetha2020ecna}. For example, metallic \sca\ is a Stoner-enhanced paramagnet and does not exhibit any magnetic ordering down to a temperature~$T$ of 0.05~K; however, stripe-type AFM fluctuations were revealed via inelastic neutron-scattering (INS) measurements~\cite{Pandey2013, Bing2019sca, Jayasekara2013}. Although the wave vector associated with the stripe AFM fluctuations in \sca\ is the same as that observed in the high-$T_{\rm c}$ $A$Fe$_2$As$_2$ parent compounds, the CoAs analogues do not exhibit SC, perhaps due to the simultaneous presence of strong ferromagnetic (FM) spin fluctuations~\cite{Wiecki2015b}; however, the INS  measurements in Ref.~\cite{Bing2019sca} indicate that dominant FM fluctuations present at high~$T$ cross over to AFM fluctuations at low~$T$\@. Electron doping of \sca\ via Ni substitution for Co results in helical AFM order with only a 1.3\% Ni substitution for Co~\cite{Sangeetha2019scna, YLi2019}. A composition-induced non-Fermi-liquid behavior associated with a magnetic quantum-critical point is also evidenced in \scna\ crystals near $x = 0.3$~\cite{Sangeetha2019scna}. No magnetic ordering down to $T=1.8$~K is found in metallic \bca\  which, however, is in proximity to a FM quantum-critical point associated with a flat electron band near the Fermi energy~$E_{\rm F}$~\cite{Sefat2009, Anand2014, Xu2013}.

\cca\ is a unique member of the 122-type CoAs family with strong magnetic frustration and itinerant magnetism~\cite{Sapkota2017}. In contrast to \sca\ and \bca\ with ucT structures, \cca\ has a cT structure with a substantial concentration of vacancies on the Co site~\cite{Quirinale2013, Anand2014Ca}. The compound undergoes A-type AFM ordering below $T_{\rm N} = 52$~K, where the in-plane Co moments are aligned ferromagnetically along the $c$~axis, while adjacent layers of Co moments are stacked antiferromagnetically.  From magnetic susceptibility $\chi$ versus temperature~$T$ measurements, the dominant interactions are FM\@. INS measurements further reveal the presence of quasi one-dimensional FM spin fluctuations in the FM square Co planes associated with a flat band near $E_{\rm F}$, as also suggested by band-structure calculations~\cite{Sapkota2017, Mao2018}.

Due to the presence of the strong magnetic frustration in \cca, it is expected that the ground state of the compound can be changed by tuning $E_{\rm F}$ via chemical substitution. Hole doping has been accomplished by substituting Fe for Co in \ccfa\ single crystals~\cite{Jayasekara2017}. The $\chi(T)$ and neutron-diffraction measurements on \ccfa\ single crystals demonstrated that the Fe substitution suppresses the A-type AFM ordering with a reduction in the ordered moment, with long-range magnetic ordering disappearing by $x = 0.12$. INS measurements on an $x = 0.15$ single crystal further show that although the same level of magnetic frustration is present in $x = 0.15$ and $x = 0$, the spin fluctuations are strongly reduced for $x = 0.15$~\cite{UelandCCFA}. Thus, hole doping eventually quenches the magnetic moments via a Stoner transition~\cite{UelandCCFA}.

In this paper, we report the growth of \ccna\ single crystals and studies of their crystallographic, magnetic, and thermal properties.  These materials are electron-doped instead of hole-doped as in \ccfa.  The $\chi(T)$ and $M$ at fixed~$H$ versus time~$t$ measurements indicate that the A-type  AFM ordering for $x=0$  is suppressed to 10.5~K for $x = 0.11$ and does not occur above 1.8~K for $x=0.16$, and that spin-glass ordering occurs below 5~K for these two compositions. Neutron-diffraction measurements of these two compositions show no evidence for long-range magnetic ordering above 4.6~K within experimental resolution.  The out-of-plane magnetic susceptibility $\chi_c$ is strongly enhanced for $x \geq 0.16$, indicating the presence of strong FM $c$-axis fluctuations which in turn suggests the presence of a nearby FM quantum-critical point (QCP). We also report density-functional theory (DFT) calculations of the competition between different magnetic structures and fluctuations.   The large anisotropy between $\chi_{ab}$ and $\chi_c$ in the composition region $0.21 < x < 0.42$ suggests quasi-one-dimensional FM spin fluctuations along the $c$~axis. The signature of critical FM spin fluctuations is also evident in the heat capacity $C_{\rm p}(T)$ for compositions $0.11 \leq x \leq 0.52$, where $C_{\rm p}(T)/T$ versus~$T$ exhibits a low-$T$ upturn having a $\log T$ dependence that is suppressed in magnetic fields applied along the $c$~axis.  From the DFT calculations, we infer the occurrence of a Stoner transition to a nonmagnetic metal at $x\approx0.5$.  For $x=1$, metallic single-crystal  \cna\ is reported to be paramagnetic down to $T = 2$~K with a small and nearly $T$-independent paramagnetic $\chi$~\cite{Cheng2012}.

The experimental details are given in Sec.~\ref{Sec:ExpDet}.  The crystallography results are presented in Sec.~\ref{Sec:Cryst}, the magnetization $M$ versus applied magnetic field $H$ isotherms and the $\chi(T)$ data in Sec.~\ref{Sec:chiM}, and the $C_{\rm p}(T)$ data in Sec.~\ref{Sec:Cp}.  The neutron-diffraction measurements are presented in Sec.~\ref{Neuts} and the DFT calculations in Sec.~\ref{Theory}.  A summary of our results including an $H$-$T$ phase diagram constructed from the $\chi(T)$ and $M(H,t)$ isotherm measurements is given in Sec.~\ref{ConcRem}.

\section{\label{Sec:ExpDet} Experimental and theoretical details}

Single crystals of \ccna\ with compositions $x$ = 0, 0.03, 0.05, 0.11, 0.16, 0.21, 0.31, 0.42, 0.52, 0.67, 0.81 and 1 were grown using a self-flux solution-growth technique. The high-purity starting materials Ca (99.999\%), Co (99.998\%), Ni (99.999\%), and As (99.9999\%) from Alfa Aesar were placed in an alumina crucible with the molar ratio \mbox{Ca:Co:Ni:As = 1.2:$4(1 - x$):$4x$:4.} The 20\% extra Ca is found to produce larger crystals. The loaded crucible was then put into a silica tube and quartz wool was placed above the crucible to extract the molten flux during centrifugation. The silica tube was then sealed under $\approx 1/4$~atm of Ar gas. The assembly was heated to 650~$^{\circ}$C at a rate of 50~$^{\circ}$C/h and held there for 6~h to avoid excess pressure created by As vapor. Then the ampule was heated to 1300~$^{\circ}$C at 60~$^{\circ}$C/h and held at that temperature for 20~h for homogenization. Finally, the tube was cooled to 1180~$^{\circ}$C at a rate of 6~$^{\circ}$C/h and the single crystals were separated from the excess flux using a centrifuge. Shiny platelike single crystals of different sizes were obtained from the growths with the $c$~axis perpendicular to the plate surfaces with typical dimensions  $3\times3\times0.5$~mm$^3$. The crystal size was found to increase with increasing Ni concentration.

A scanning-electron microscope from JEOL equipped with an energy-dispersive x-ray spectroscopy (EDS) attachment was used to check the phase homogeneity and average compositions of the crystals. The EDS scans were taken at multiple points on both sides of the crystals as well as on cleaved crystals to confirm the homogeneity and the absence of any secondary phases.

Single-crystal x-ray diffraction (XRD) measurements were performed at room temperature on a Bruker D8 Venture diffractometer operating at 50~kV and 1~mA equipped with a Photon 100 CMOS detector, a flat graphite monochromator and a Mo~K$\alpha$ I$\mu$S microfocus source ($\lambda = 0.71073$~\AA). The raw frame data were collected using the Bruker APEX3 software package~\cite{APEX2015}, while the frames were integrated with the Bruker SAINT program~\cite{SAINT2015} using a narrow-frame algorithm for  integration of the data and were corrected for absorption effects using the multiscan method (SADABS)~\cite{Krause2015}.  The atomic thermal factors were refined anisotropically.  Initial models of the crystal structures were first obtained with the program SHELXT-2014 \cite{Sheldrick2015A} and refined using the program SHELXL-2014 \cite{Sheldrick2015C} within the APEX3 software package.

The temperature $T$- and magnetic field $H$-dependent magnetic measurements were carried out using a Quantum Design, Inc.,  magnetic-properties measurement system (MPMS) SQUID magnetometer in the range $T=1.8$ to~300~K with $H$ up to 5.5~T (1~T~$\equiv 10^4$~Oe). The heat capacity $C_{\rm p}(H,T)$ measurements were performed using the relaxation technique in a Quantum Design, Inc., Physical-Properties Measurement System (PPMS) in the ranges $T=1.8$--300~K and \mbox{$H = 0$--9~T.}

Single-crystal neutron-diffraction experiments were performed in $H=0$ using the TRIAX triple-axis spectrometer at the University of Missouri Research Reactor (MURR). An incident neutron beam of energy \mbox{$E_i = 30.5$~meV} (wavelength $\lambda= 1.6377$ {\AA}) was directed at the sample using a pyrolytic graphite (PG) monochromator. The $\lambda/2$ component  present in the beam was removed using PG filters placed before the monochromator and between the sample and analyzer. Beam divergence was limited using collimators before the monochromator; between the monochromator and sample; sample and analyzer; and analyzer and detector of $60^\prime-60^\prime-40^\prime-40^\prime$, respectively.  Crystals of mass 20~mg were mounted on the cold tip of an Advanced Research Systems closed-cycle refrigerator with a base temperature in the 4.1 to 4.8~K range. The crystals were aligned in the $(H00)$ and $(00L)$ plane. Extended diffraction patterns along principal directions confirm the $I4/mmm$ tetragonal symmetry with lattice constants $a = 3.981(3)$ and {$c = 10.07(4)$~\AA} at 4.8 K for the $x=0.16$ crystal and $a = 3.980(3)$ and {$c = 10.10(4)$~\AA}  for the $x=0.11$ crystal.

DFT calculations were performed using a full-potential linear-augmented-plane-wave (FP-LAPW) method, as implemented in \textsc{wien2k}~\cite{wien2k}.
We used the virtual-crystal approximation (VCA) to simulate the doping effects.
The generalized-gradient approximation (GGA) of Perdew, Burke, and Ernzerhof~\cite{perdew1996} was used for the correlation and exchange potentials.
To generate the self-consistent potential and charge, we employed $R_\text{MT}\cdot K_\text{max}=8.0$ with Muffin-tin (MT) radii $R_\text{MT}= 2.7$, 2.4, and 2.5 a.u.\ for Ca, Co, and As, respectively.
The calculations were performed with 264 $k$-points in the irreducible Brillouin zone (BZ).
They were iterated until the charge differences between consecutive iterations was smaller than $1\times10^{-3}\,e$ and the total energy difference less than 0.01~mRy.

We also calculated the magnetocrystalline anisotropy energy (MAE), which originates from the spin-orbit coupling (SOC)~\cite{ke2015prb}. The MAE is calculated as \mbox{$K=E_{100}{-}E_{001}$}, where $E_{001}$ and $E_{100}$ are the total energies for the magnetization oriented along the [001] and [100] directions, respectively.  Positive (negative) $K$ corresponds to easy-$c$-axis (easy-$ab$-plane) anisotropy.  The SOC is included using the second-variation procedure~\cite{li1990prb}.

The composition-dependent experimental lattice parameters listed in Table~\ref{CrystalData} in the following section were adopted in the calculations.

\section{\label{Sec:Cryst} Crystallography}

\begin{figure}
\includegraphics[width = 1.5in]{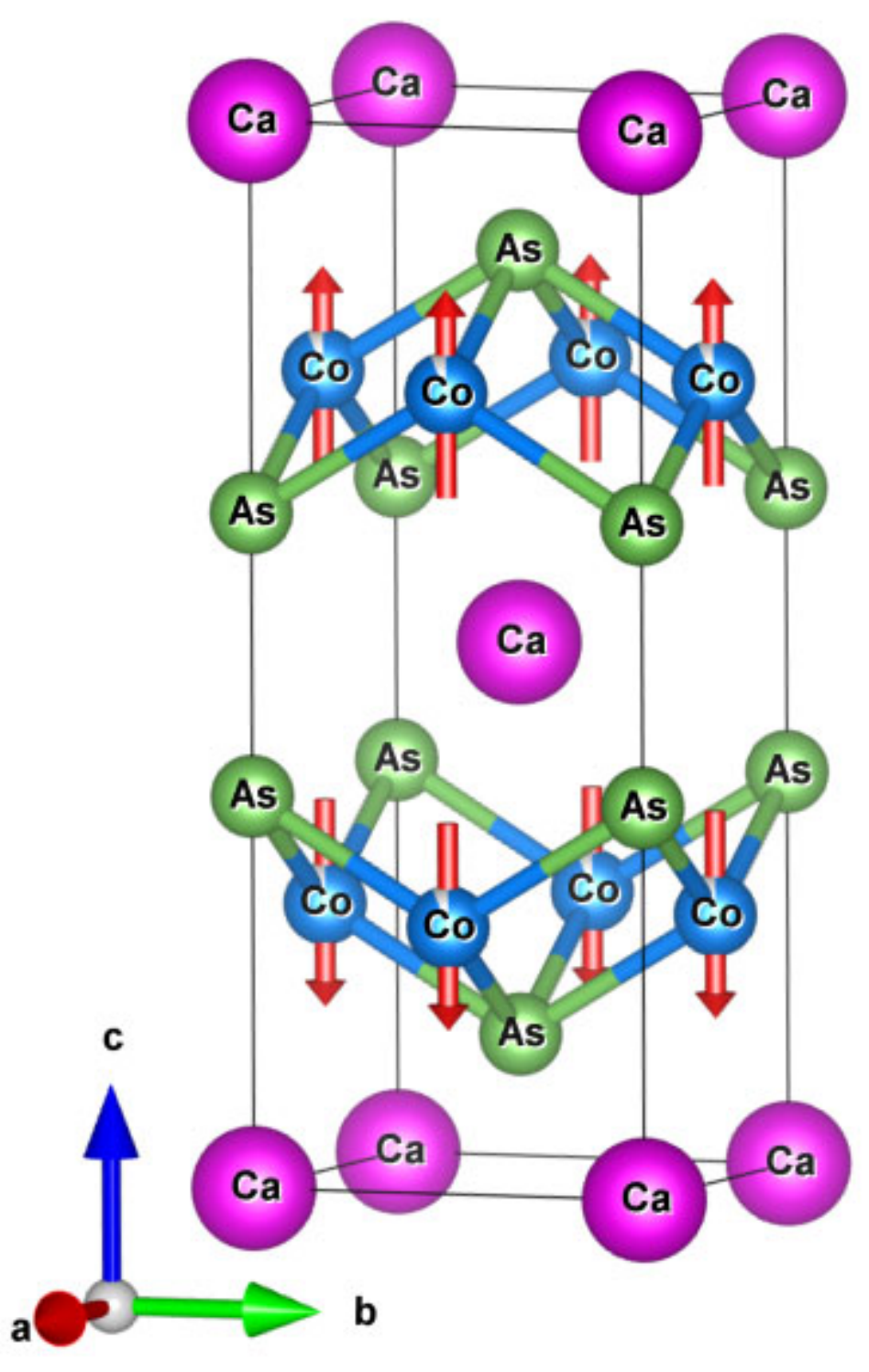}
\caption{Chemical and A-type AFM structure of \cca. The ThCr$_2$Si$_2$-type chemical unit cell is body-centered tetragonal and the AFM unit cell is A-type with the ordered moments aligned along the $c$~axis.  The crystal structure was drawn using the VESTA program~\cite{Momma2011}.}
\label{Crystalstructure}
\end{figure}

The body-centered-tetragonal crystal structure of \cca\ is shown in Fig.~\ref{Crystalstructure}~\cite{Momma2011}.  The room-temperature crystallographic data for \ccna\ crystals obtained from single-crystal XRD measurements are listed in Table~\ref{CrystalData}. The listed compositions are the average compositions of the crystals from EDS measurements. The single-crystal XRD refinement does not allow  simultaneous refinement of the total occupation of the transition metal site and the fraction of Co/Ni at that site. Thus, only the total occupancies were refined based on the Co/Ni ratio taken from the EDS data. All the compositions including the parents \cca\ and \cna\ crystallize in the cT \tcs-type structure.

\begin{figure}
\includegraphics[width = 3.3in]{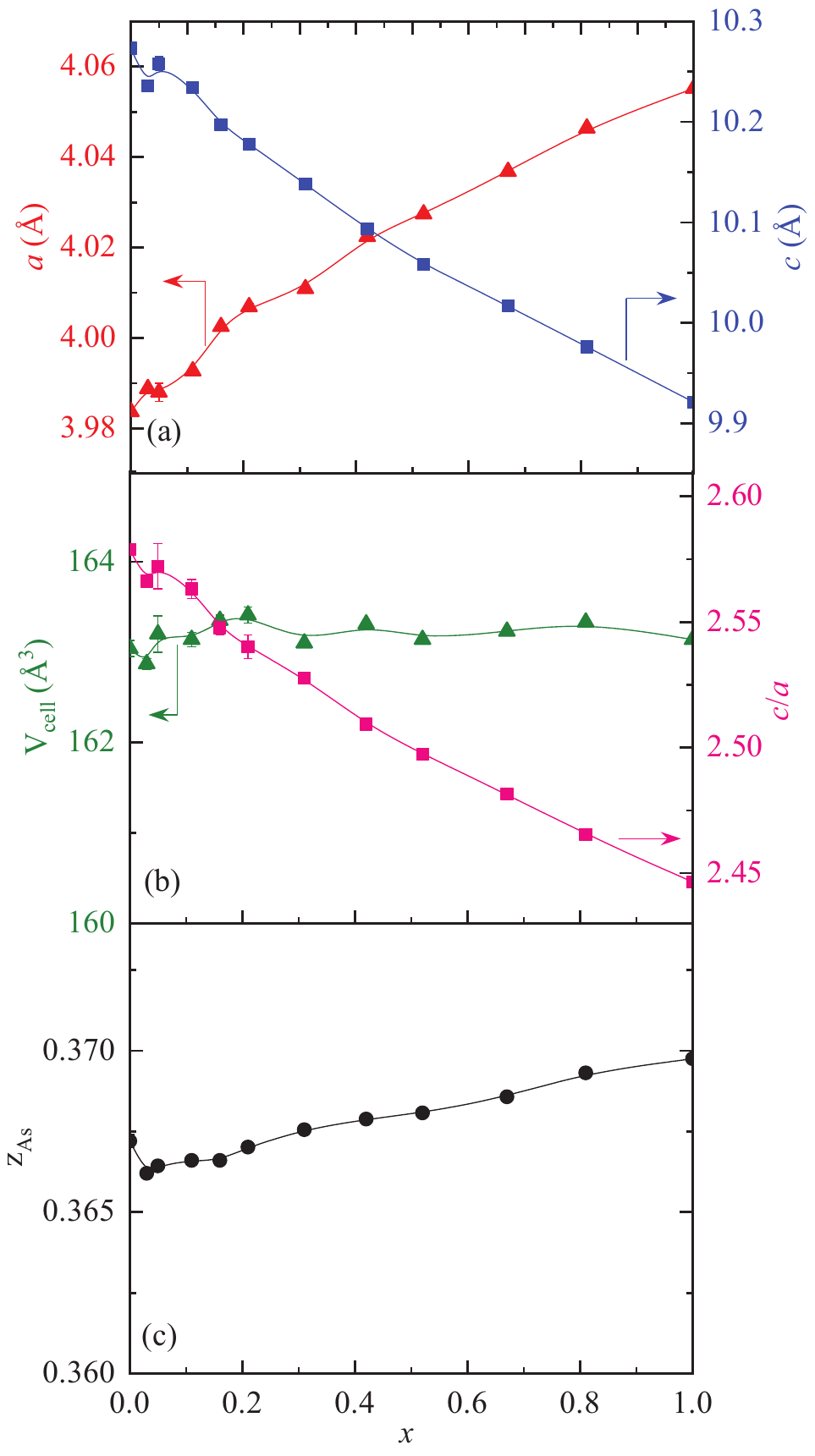}
\caption{Unit cell parameters (a)~tetragonal lattice parameters $a$ and $c$, (b)~Unit cell volume $V_{\rm cell}$ and $c/a$ ratio, and \mbox{(c)~$c$-axis} As position parameter $z_{\rm As}$ obtained from the single-crystal XRD measurements on \ccna\ crystals as a function of Ni substitution~$x$.  The solid lines are guides to the eye.}
\label{Fig_Cell_parameters}
\end{figure}

The tetragonal lattice parameters $a$ and $c$, unit-cell volume $V_{\rm cell}$, $c/a$ ratio, and the $z$-axis As position parameter $z_{\rm As}$ are plotted as a function of the Ni concentration $x$  in Fig.~\ref{Fig_Cell_parameters}. The $a$ lattice parameter increases and the $c$ lattice parameter decreases nonlinearly with $x$.  As a result the unit-cell volume is almost constant throughout the composition range. Although all compositions form in the cT structure, the decrease in the ratio $c/a$ with increasing $x$ may result in a significant change in the interplanar magnetic coupling between the Co spins. The $z_{\rm As}$ parameter initially decreases from $x=0$ to $x = 0.03$, but then increases with further increases in Ni concentration. The EDS scans taken at points on both sides of each crystal confirm the homogenous nature of the crystals. The average compositions estimated by EDS indicate that the vacancy concentration on the transition-metal site decreases with increasing Ni substitution and that the stoichiometric composition ${\rm CaNi_2As_2}$ is obtained within the error bar for $x=1$. 

\begin{table*}
\caption{\label{CrystalData} Tetragonal lattice parameters $a$ and~$c$, $c/a$ ratio, unit-cell volume $V_{\rm cell}$, and fractional $c$-axis position of the As site $z_{\rm As}$ obtained for \ccna\ from room-temperature single-crystal XRD measurements. The listed values of~$x$ are the average values obtained from EDS analyses.}
\begin{ruledtabular}
\begin{tabular}{ cccccc }
 Compound  & $a$ (\AA)  & $c$ (\AA) & $c/a$ & $V_{\rm cell}$ (\AA$^3$) &  $z_{\rm As}$\\
\hline
CaCo$_{1.86(2)}$As$_2$                               	&   3.9837(4)   &   10.2733(4)   & 2.5788(6)   &   163.04(9)   &     0.3672(4)   \\
Ca(Co$_{0.970(3)}$Ni$_{0.030(3)}$)$_{1.87(4)}$As$_2$    &   3.9888(6)    &   10.236(2)    & 2.5661(2)   &   162.87(6)  &     0.36620(5)   \\
Ca(Co$_{0.95(1)}$Ni$_{0.05(1)}$)$_{1.86(4)}$As$_2$    &   3.988(2)    &   10.258(7)    & 2.5722(4)   &  163.2(2)  &      0.36643(8)   \\
Ca(Co$_{0.89(1)}$Ni$_{0.11(1)}$)$_{1.87(4)}$As$_2$      &   3.9927(8)    &   10.234(3)    &  2.5631(3)   &  163.14(8)  &     0.3666(1)   \\
Ca(Co$_{0.84(1)}$Ni$_{0.16(1)}$)$_{1.89(6)}$As$_2$      &   4.0025(6)    &   10.197(2)    &  2.5476(1)   & 163.35(6)  &      0.3666(1)   \\
Ca(Co$_{0.79(1)}$Ni$_{0.21(1)}$)$_{1.86(3)}$As$_2$      &   4.0069(8)    &   10.178(4)    &  2.5401(6)   & 163.41(9)  &      0.36701(9)   \\
Ca(Co$_{0.69(1)}$Ni$_{0.31(1)}$)$_{1.87(3)}$As$_2$      &   4.0109(1)    &   10.1382(5)    &  2.5276(1)   &  163.10(1)  &     0.36755(8)   \\
Ca(Co$_{0.58(1)}$Ni$_{0.42(1)}$)$_{1.87(3)}$As$_2$      &   4.0224(3)    &   10.0935(1)    &   2.5093(2)   & 163.31(3)  &     0.36788(5)   \\
Ca(Co$_{0.48(1)}$Ni$_{0.52(1)}$)$_{1.87(4)}$As$_2$      &   4.0274(3)    &   10.058(1)    &   2.4973(5)   &  163.14(3)  &    0.36807(4)   \\
Ca(Co$_{0.33(1)}$Ni$_{0.67(1)}$)$_{1.89(4)}$As$_2$      &   4.0368(3)    &  10.017(1)     &  2.4814(4)   &  163.23(3)    &  0.36857(6)    \\
Ca(Co$_{0.19(1)}$Ni$_{0.81(1)}$)$_{1.94(3)}$As$_2$      &  4.0464(2)     &  9.976(1)     &  2.4654(3)   &  163.33(2)    &  0.36931(7)    \\
CaNi$_{1.97(4)}$As$_2$      &   4.0552(2)    &  9.921(1)     &  2.4464(4)   &  163.14(2)    &  0.36975(4)    \\
\end{tabular}
\end{ruledtabular}
\end{table*}

\section{\label{Sec:chiM} Magnetic measurements}

\subsection{Magnetic susceptibility}

\begin{figure*}
\includegraphics[width = \textwidth]{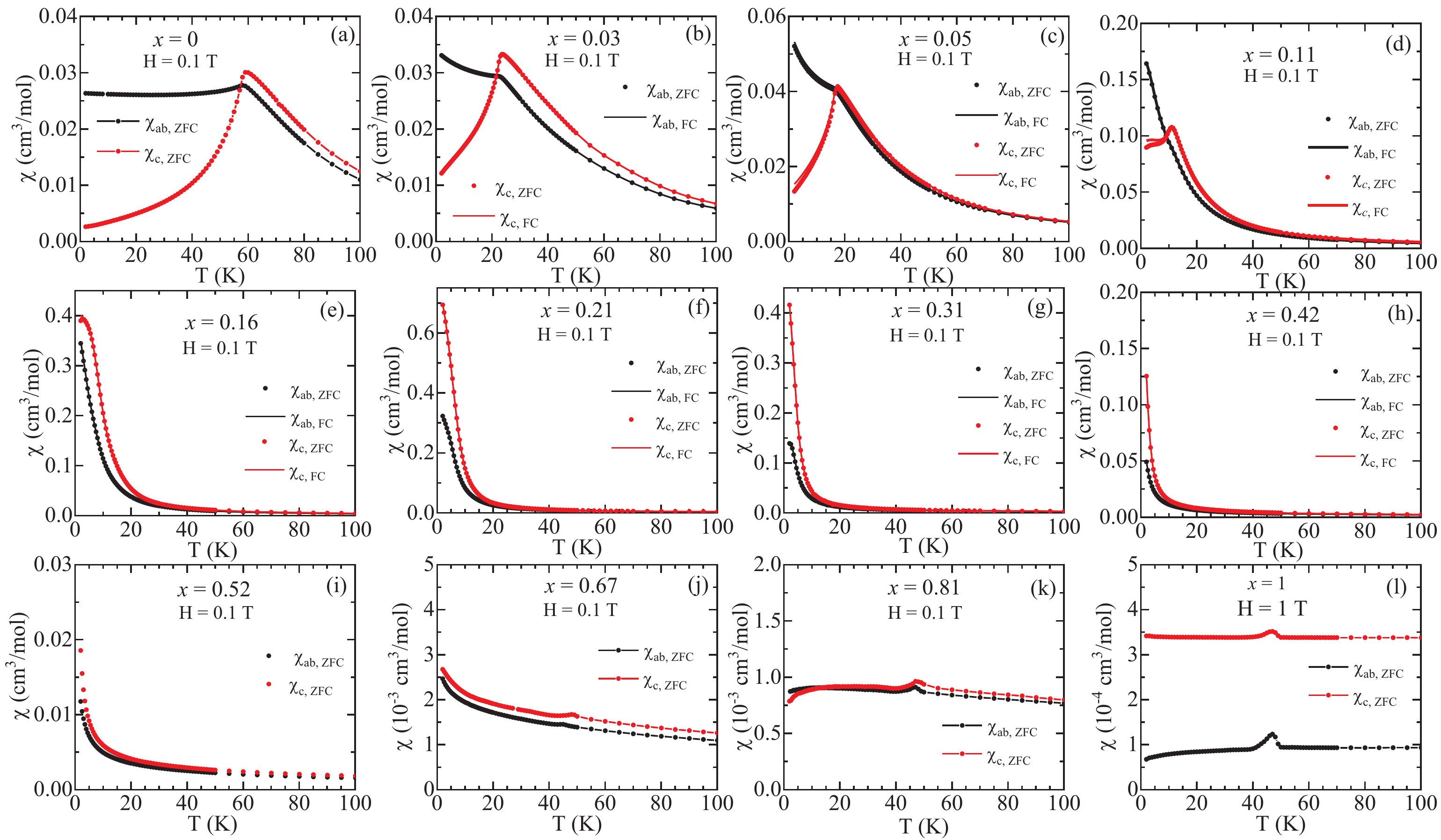}
\caption{The temperature $T$ dependence of magnetic susceptibility $\chi$ of \ccna\ crystals with compositions $x$ in applied magnetic fields~$H$ in the $ab$ plane and along the $c$ axis as indicated. The data for $x=0.11$ and~0.16 show a transition in $\chi_c(T)$ but not in $\chi_{ab}(T)$ as discussed separately in Sec.~\ref{subsec:LowTanomaly}.  Note the large decrease with~$x$ of the ordinate scales for $\chi$, indicating the loss of Co magnetic character for $x\gtrsim0.5$.  The small anomalies at $\approx 50$~K in panels (j), (k), and~(l) may arise from small amounts of O$_2$ adsorbed on the crystal surfaces~\cite{Freiman2004}. }
\label{Fig_M-T_all_separate}
\end{figure*}

\begin{figure}
\includegraphics[width = 2.5in]{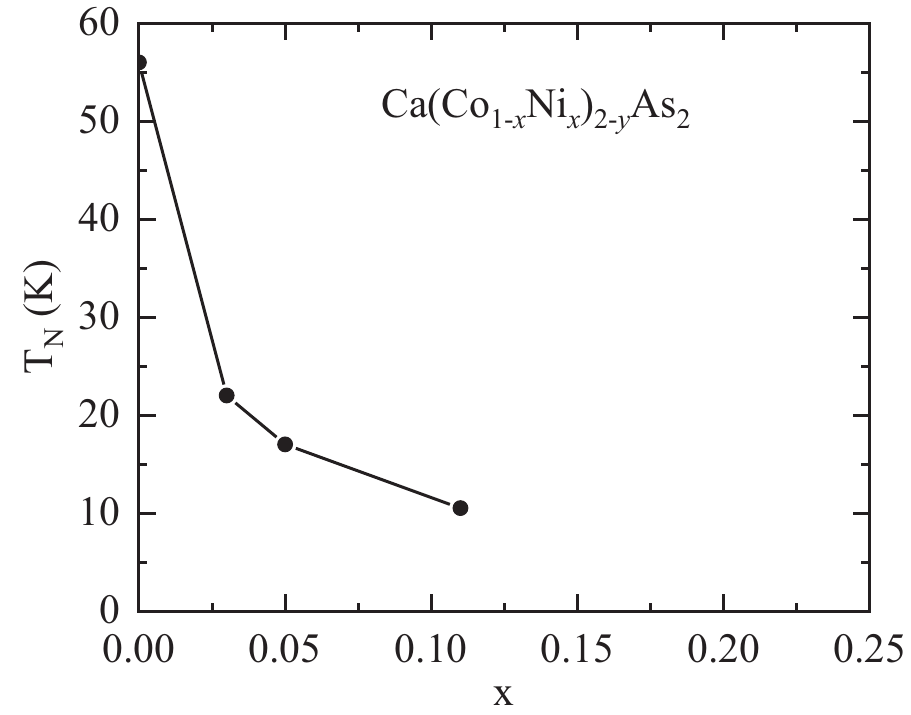}
\caption{Dependence of the N\'eel temperature $T_{\rm N}$ of \ccna\ crystals on composition~$x$.}
\label{Ca_CoNi2As2_TNvsx}
\end{figure}

The $T$-dependent magnetic susceptibilities $\chi = M/H$ for \ccna\ crystals with $x$ = 0, 0.03, 0.05, 0.11, 0.16, 0.21, 0.31, 0.42, 0.52, 0.67, 0.81, and 1 are shown in Figs.~\ref{Fig_M-T_all_separate}(a)--\ref{Fig_M-T_all_separate}(l), respectively, for $H \parallel ab$ and $H \parallel c$ under both zero-field-cooled (ZFC) and field-cooled (FC) conditions as noted in the respective panels. The parent compound \cca\ shows A-type AFM order with the ordered moments aligned along the $c$~axis below the N\'eel temperature $T_{\rm N} = 52$~K~\cite{Anand2014Ca}. The $c$-axis susceptibility $\chi_c$ approaches zero for $T \rightarrow 0$ and the in-plane susceptibility $\chi_{ab}$ is almost independent of $T$ below $T_{\rm N}$, consistent with collinear AFM order along the $c$~axis according to molecular-field theory~\cite{Johnston2015}.  The $T_{\rm N}$ is determined from the $\chi_{ab}(T)$ data using the Fisher criterion~\cite{Fisher1962} where here the temperature of the peak in $d(\chi_{ab} T)/dT$ corresponds to $T_{\rm N}$\@.  The $T_{\rm N}$ drops from 52 K for $x=0$ to 22~K with only 3\% Ni doping and then decreases further (see Table~\ref{Tab.chidata} below) as shown in Fig.~\ref{Ca_CoNi2As2_TNvsx}.

\begin{figure*}
\includegraphics[width = 5in]{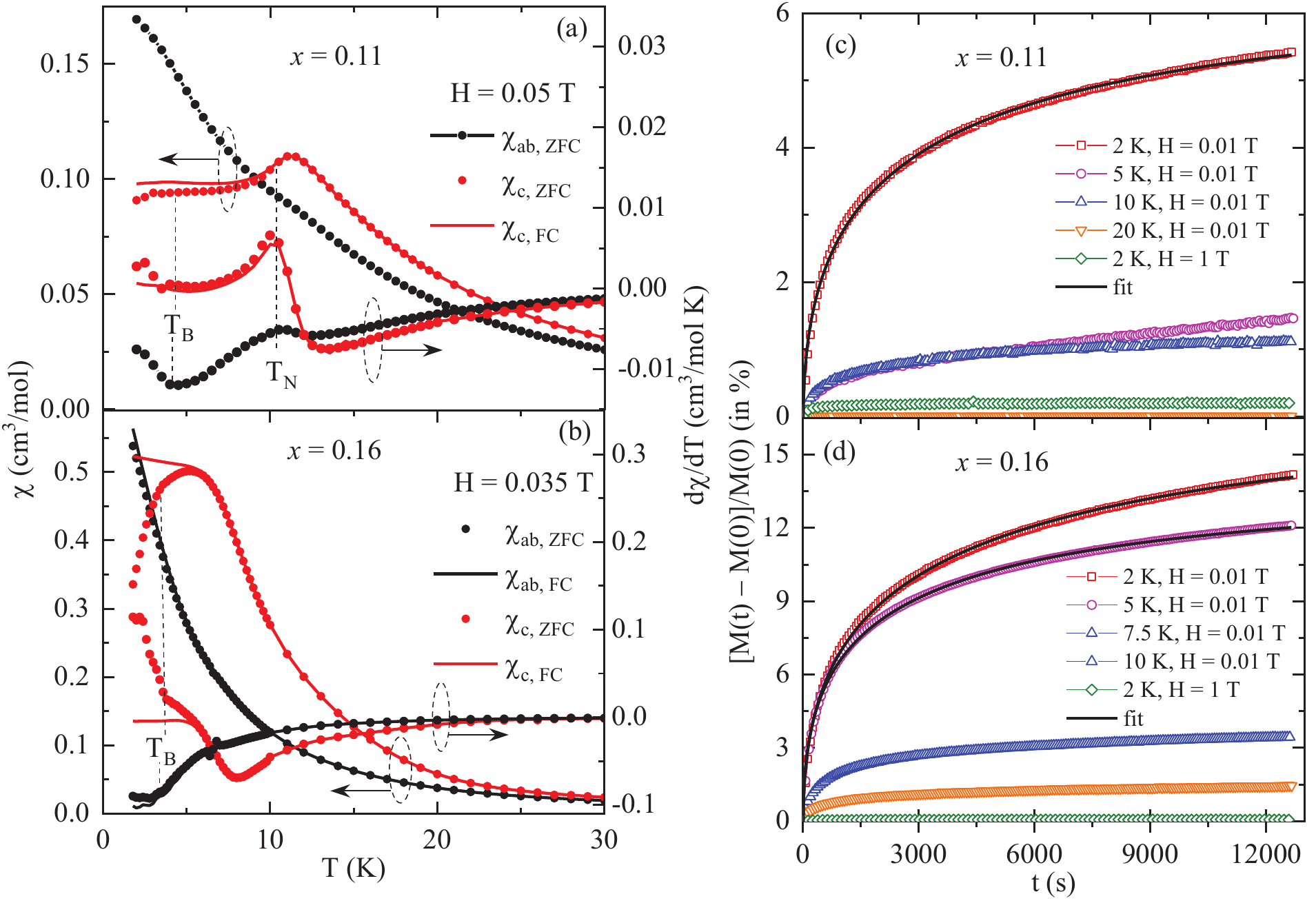}
\caption{Measurements of $\chi_{ab}(T)$ and $\chi_c(T)$ below 30~K for (a)~$x = 0.11$ and (b)~$x=0.16$ crystals measured in $H = 0.05$ and 0.035~T, respectively, under both ZFC and FC conditions (left ordinates) along with the temperature derivative of the susceptibilities (right ordinates).  Clear signatures of two transitions at $T_{\rm N}$ and $T_{\rm B}$ are observed for $x=0.11$. The time~$t$ dependence of relative magnetic relaxation $[M(t) - M(0)]/M(0)$ measured at different temperatures with $H\parallel c$ is plotted in (c) and (d) for $x = 0.11$ and 0.16, respectively. The solid black lines for $T=2$~K ($x=0.11$) and $T=2$ and 5~K ($x=0.16$) with $H=0.01$~T are fits of the respective relaxation data by Eq.~(\ref{Eq:Mrelax}). }
\label{Fig_susceptibility_11-16}
\end{figure*}

With increasing~$x$, the values of $\chi_{ab}$ and $\chi_{c}$ both increase in the ordered state, indicating enhancements by FM  fluctuations. For 5\% Ni doping a small hysteresis between the ZFC and FC data for $\chi_{c}$ in Fig.~\ref{Fig_M-T_all_separate}(c) occurs at $T \ll T_{\rm N}$. The thermal hysteresis in $\chi_{c}(T)$ increases for $x = 0.11$, where along with an AFM transition at $T_{\rm N} = 10.5$~K, another anomaly is observed at $T_{\rm B} \approx 4.8$~K which is clearly visible in the lower fields as shown in Fig.~\ref{Fig_susceptibility_11-16}(a) for $H = 0.05$~T\@. For $x = 0.16$, the signature of an AFM transition for $H = 0.1$~T in Fig.~\ref{Fig_M-T_all_separate}(e) has almost disappeared, replaced by a strong FM-like increase, although the low-field data in $H = 0.05$~T in Fig.~\ref{Fig_susceptibility_11-16}(b) confirm the presence of two magnetic transitions at $T_{\rm N} \sim 6.5$~K and $T_{\rm B} \sim 3.5$~K, where $T_{\rm B}$ is identified below as a blocking temperature. A separate discussion of the low-$T$ anomaly observed at $T_{\rm B}$ for $x = 0.11$ and 0.16 is given in the following section. No signature of AFM ordering of any kind is observed for $x > 0.16$.

For $x>0.16$, Fig.~\ref{Fig_M-T_all_separate} shows that $\chi_{c}$ strongly increases and becomes much larger than $\chi_{ab}$.  At the lowest measured $T= 2$~K, $\chi_{c} \approx 2\chi_{ab}$ for $x=0.21$. The ratio $\chi_{c}/\chi_{ab}$ increases further for $x = 0.31$. These observations indicate a strong enhancement of FM $c$-axis spin fluctuations for $x = 0.21$ and 0.31. The FM spin fluctuations and the anisotropy is found to persist up to $x = 0.52$. The suppression of AFM order associated with the development of strong FM fluctuations in \ccna\ crystals suggests the presence of a nearby FM quantum-critical point (QCP) close to $x = 0.16$. We note that a FM QCP can be avoided if preempted by an AFM transition~\cite{Brando2008, Lengyel2015, Hamann2019}.

The $\chi(T)$ data for the $x = 0.67$ and 0.81 crystals in Fig.~\ref{Fig_M-T_all_separate} are significantly reduced in magnitude and are almost independent of $T$ with a small anisotropy.  The $\chi$ data are also almost independent of $T$ for $x = 1$, and exhibit larger anisotropy than for $x=0.67$ and 0.81. A  composition-independent anomaly is observed at $T \sim 50$~K for all three compositions that may be associated with a small amount of O$_2$ adsorbed on the crystal surfaces~\cite{Freiman2004}.

\subsection{\label{subsec:LowTanomaly} Low-temperature magnetic features for $x = $ 0.11 and 0.16}

\begin{figure*}[ht!]
\includegraphics[width = 6.in]{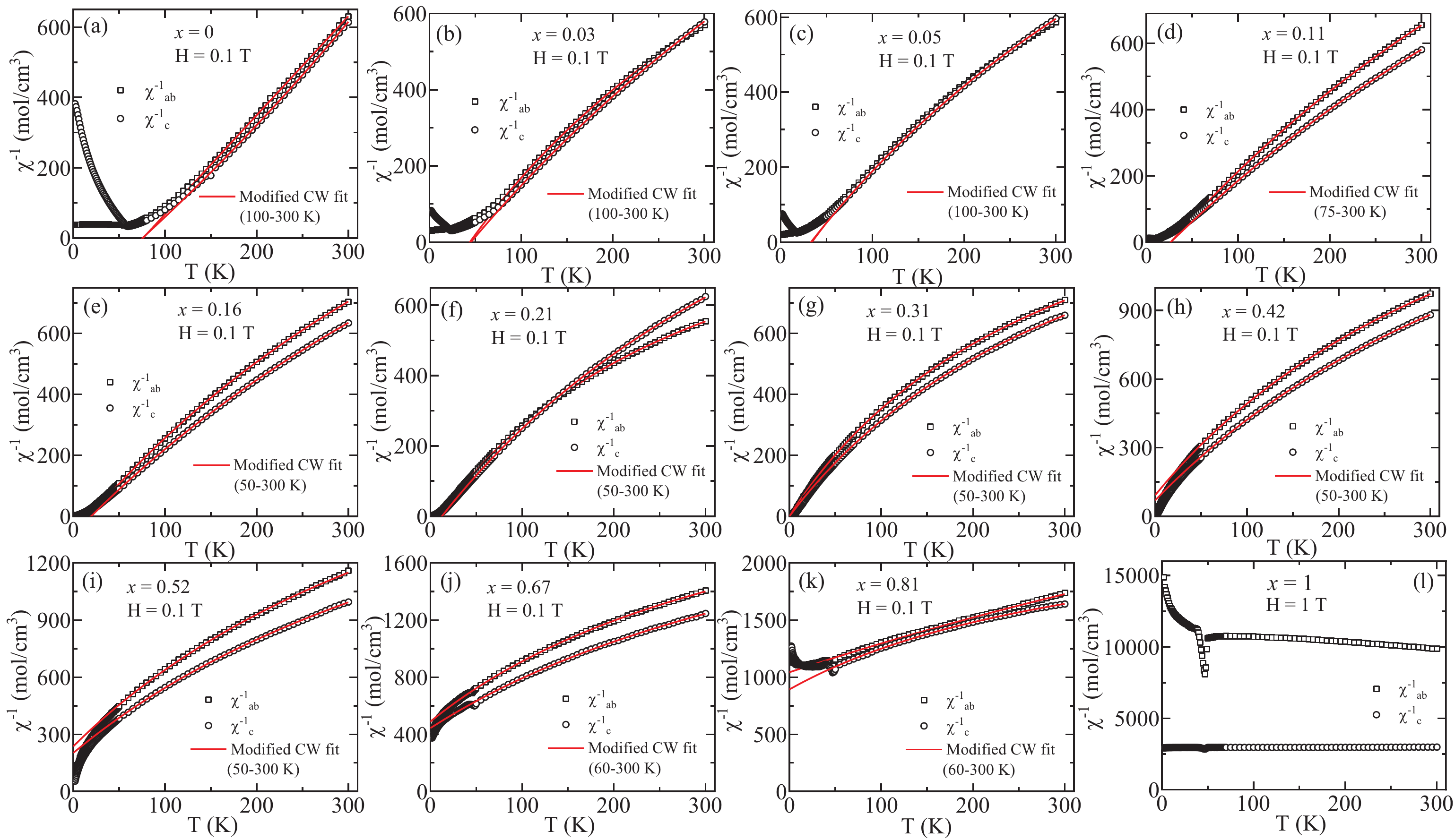}
\caption{Temperature $T$ dependence of inverse magnetic susceptibility $\chi^{-1}$ of \ccna\ crystals with $x = 0$--1 in panels (a)--(h) for $H \parallel ab$ and $H \parallel c$ as indicated. The anomalies at $\approx 50$~K in (j)--(l) are likely experimental artifacts arising from AFM ordering of small amounts of O$_2$ condensed on the crystals.  The red solid lines are fits by the modified Curie-Weiss law in Eq.~(\ref{Eq.ModCurieWeiss}) over the temperature range noted in the respective panel. The positive values of $\chi_0$ are responsible for the negative curvature in the fits.  The overall decrease in the magnitude  and associated weak $T$ dependence of $\chi(T)$ are responsible for the increasingly negative unphysical values of $\theta_{\rm p\alpha}$ listed in Table~\ref{Tab.chidata} for $x=0.42$--0.81.}
\label{Fig_Inversse_susceptibility}
\end{figure*}

The anomalies in $\chi(T)$ at low~$T$ observed in Fig.~\ref{Fig_M-T_all_separate} for the $x = 0.11$ and 0.16 crystals were further investigated by measuring $\chi(T)$ at smaller $H$\@. The respective $\chi(T)$ data measured in $H = 0.05$~T and 0.035~T for $x = 0.11$ and 0.16 are shown in Figs.~\ref{Fig_susceptibility_11-16}(a) and \ref{Fig_susceptibility_11-16}(b), respectively, along with the respective $T$  derivatives. For $x = 0.11$, $T_{\rm N} = 10.5$~K\@. The ZFC and FC $\chi_c$ data for each composition diverge from each other below $T_{\rm N}$, exhibiting an anomaly at a blocking temperature $T_{\rm B} = 4.8$~K\@. The FM fluctuations considerably increase for $x = 0.16$ and it is difficult to ascertain whether long-range AFM order occurs for this composition. Indeed, the zero-field neutron diffraction measurements discussed below show no evidence for long-range AFM ordering for $x = 0.16$.  On the other hand, the $\chi(T)$ data for $x=0.16$ show clear evidence for a blocking temperature $T_{\rm B} \approx 3.5$~K\@.

Thus in \ccna\ crystals the AFM order is suppressed and FM fluctuations increase significantly with increasing Ni substitution.  It is therefore likely that for the $x = 0.11$ and~0.16 compositions AFM and FM interactions compete with each other at low~$T$ which gives rise to a metastable glassy state at a temperature $T_{\rm B}$, similar to that observed in many reentrant AFM spin-glass systems~\cite{Dho2002,Pakhira2016,Viswanathan2009}. To further elucidate this scenario, we measured the magnetic relaxation behavior at different temperatures for both compositions. The crystals were zero-field-cooled from the paramagnetic (PM) state to the measured temperature at which a magnetic field was switched on and the time evolution of the magnetization $M(t)$ was recorded. As seen from Figs.~\ref{Fig_susceptibility_11-16}(c) and \ref{Fig_susceptibility_11-16}(d), a clear $t$ dependence of the relative change in magnetization $M(t)$ is observed for $T \lesssim T_{\rm B}$ in the $x = 0.11$ and~0.16 crystals. It is interesting to see that although the relaxation for $T_{\rm B} < T < T_{\rm N}$\@ is negligible for $x = 0.11$, the relaxation is quite significant at $T = 2$ and 5~K for $x = 0.16$. This difference suggests that the strong enhancement of FM fluctuations along the $c$ axis and resultant strong competition between the FM and AFM interactions significantly weakens AFM ordering and pushes the system towards metastability at low~$T$\@. This is consistent with the observation of thermomagnetic irreversibility just below $T_{\rm N}$ for $x=0.11$. Moreover, magnetic relaxation is observed in Figs.~\ref{Fig_susceptibility_11-16}(c,d)  at $T = 2$~K for small $H = 0.01$~T, whereas the larger $H= 1$~T is seen to prevent the metastable state from forming for both $x=0.11$ and $x=0.16$.

\begin{table}
\caption{\label{Tab.Relaxation} Parameters obtained by fitting the magnetic relaxation behavior of \ccna\ crystals with \mbox{$x = 0.11$} and 0.16 in Figs.~\ref{Fig_susceptibility_11-16}(c) and \ref{Fig_susceptibility_11-16}(d) by the stretched-exponential function in Eq.~(\ref{Eq:Mrelax}) for $H = 0.01$~T applied along the $c$~axis.}
\begin{ruledtabular}
\begin{tabular}{cccc}	
$x$ 									&		 $T$~(K)			& $\tau~(10^3$ s)	 &  $\alpha$  \\
\hline
0.11	& 2	& 2.98(8)	 &  0.46(1) \\
0.16	& 2	& 3.6(1)	 &  0.45(1) \\
        & 5	& 2.27(5)	 &  0.44(1) \\
\end{tabular}
\end{ruledtabular}
\end{table}

The relaxation behavior observed in Figs.~\ref{Fig_susceptibility_11-16}(c,d) for $T \lesssim T_{\rm B}$ is described well by the stretched-exponential dependence
\bea
\frac{M(t)}{M(t=0)}  =1- e^{-(t/\tau)^\alpha},
\label{Eq:Mrelax}
\eea
similar to that observed in spin-glass systems~\cite{Mydosh1993, Johnston2006}.  The fits at low temperatures are shown as the solid black curves in Figs.~\ref{Fig_susceptibility_11-16}(c,d) and the fitted parameters $\tau$ and~$\alpha$ are listed in Table~\ref{Tab.Relaxation}. Although the parameter $\alpha$ is similar for both crystals, a significantly larger $\tau$ is observed at $T=2$~K for $x = 0.16$. This suggests that the competition between AFM and FM interactions is stronger in the \mbox{$x = 0.16$} crystal, consistent with the $\chi(T)$ data in Fig.~\ref{Fig_M-T_all_separate}.  Physical interpretations of the parameters $\tau$ and~$\alpha$ are given in Ref.~\cite{Johnston2006}.

\subsection{Magnetic susceptibility}

\begin{table*}[ht!]
\caption{\label{Tab.chidata} Parameters obtained from the modified Curie-Weiss fits to $\chi^{-1}(T)$ data in the respective temperature range noted in Fig.~\ref{Fig_Inversse_susceptibility} for \ccna\ crystals with $x=0$--0.81 using Eq.~(\ref{Eq.ModCurieWeiss}). The  parameters listed are the N\'eel temperature $T_{\rm N}$ and blocking temperature $T_{\rm B}$, the magnetic-field direction~$\alpha$, the $T$-independent contribution to the magnetic susceptibility $\chi_{0\alpha}$,  the Curie constant per mol $C_\alpha$, the effective moment per transition metal atom $\mu{\rm_{eff\alpha}}$ calculated using Eq.~(\ref{Eq.mueff}), the spherical average $\mu_{\rm eff, ave} =\sqrt{ (2\mu_{{\rm eff}\,ab}^2 + \mu_{{\rm eff}\,c}^2)/3}$, the Weiss temperature $\theta\rm_{p\alpha}$, and the spherical average $\theta_{\rm p, ave} = \frac{2}{3}\theta_{ab} + \frac{1}{3}\theta_{c}$. }
\begin{ruledtabular}
\begin{tabular}{ccccccccc}	
  	Compound		& $T_{\rm N}$, $T_{\rm B}$  							& Field			& $\chi_{0\alpha}$ 	& $C_{\alpha}$ &  $\mu_{\rm eff\alpha}$ & $\mu_{\rm eff, ave}$ 	& $\theta_{\rm p\alpha}$ & $\theta_{\rm p, ave}$  \\
 					&			& direction $\alpha$		& $\rm{\left(10^{-4}~\frac{cm^3}{mol}\right)}$	 & $\rm{\left(\frac{cm^3 K}{mol}\right)}$    & $\rm{\left(\frac{\mu_B}{Co+Ni}\right)}$& $\rm{\left(\frac{\mu_B}{Co+Ni}\right)}$  & (K) & (K)\\
\hline

CaCo$_{1.86(2)}$As$_2$                         & 	52, --- 	& $H\parallel ab$ 	& 0.03(2) 	& 0.443(3) 	& 1.37(1)		& 1.39(1) 	&    59.5(9) & 61(1)   \\
						                       &        & $H\parallel c$ 	& -0.2(2) 	& 0.480(1) 	& 1.43(1) 		&	      	&    64.5(3) &		    \\
Ca(Co$_{0.97}$Ni$_{0.03}$)$_{1.86}$As$_2$      & 	22, --- 	& $H\parallel ab$ 	& 5.84(5) 	& 0.299(1) 	& 1.14(1)	& 1.17(1) &    44.2(3) & 44.6(5)	\\
						                       &        & $H\parallel c$ 	& 3.52(3) 	& 0.349(1) 	& 1.22(1) 		&	      &    45.4(2) &		    \\
Ca(Co$_{0.95}$Ni$_{0.05}$)$_{1.86}$As$_2$      & 	17, --- 	& $H\parallel ab$ 	& 5.80(4) 	& 0.298(1) 	& 1.13(1)		& 1.14(1) &    34.1(3) & 33.7(6)	\\
						                       &        & $H\parallel c$ 	& 4.70(4) 	& 0.322(1) 	& 1.17(1) 		&	      &    33.1(3) &		    \\
Ca(Co$_{0.89}$Ni$_{0.11}$)$_{1.87}$As$_2$   & 10.5, 4.8 & $H\parallel ab$ 	& 3.66(3) 	& 0.318(1) 	& 1.16(1)		& 1.19(1) &    26.7(2) & 26.7(4)	\\
						                       &        & $H\parallel c$ 	& 3.71(2) 	& 0.369(1) 	& 1.25(1) 		&	      &    26.8(2) &		    \\
Ca(Co$_{0.84}$Ni$_{0.16}$)$_{1.89}$As$_2$  & ---, 3.5& $H\parallel ab$ 	&  4.06(2) 	&  0.285(1) & 1.09(1)		& 1.12(1) &    18.4(1) & 18.7(2)	\\
						                       &        & $H\parallel c$ 	&   3.79(2)	&  0.335(1) & 1.18(1)		&	      &    19.5(1) &		    \\
Ca(Co$_{0.79}$Ni$_{0.21}$)$_{1.86}$As$_2$      &     	& $H\parallel ab$ 	&  6.67(3) 	&  0.268(1) & 1.07(1)		& 1.09(1) &    10.9(2) & 11.4(4)	\\
						                       &        & $H\parallel c$ 	&   5.52(3)	&  0.303(1) & 1.13(1)		&	      &    12.6(2) &		    \\
Ca(Co$_{0.69}$Ni$_{0.31}$)$_{1.87}$As$_2$      &     	& $H\parallel ab$ 	&  6.98(5) 	&  0.214(2) & 0.95(1)		& 0.98(1) &    $-0.3(4)$ & $-0.2(7)$	\\
						                       &        & $H\parallel c$ 	&   6.77(4)	&  0.252(1) & 1.03(1)		&	      &    $-0.1(3)$ &		    \\
Ca(Co$_{0.58}$Ni$_{0.42}$)$_{1.87}$As$_2$      &     	& $H\parallel ab$ 	&  4.36(3) 	&  0.188(1) & 0.89(1)		& 0.92(1) &    $-18.2(5)$ &       	\\
						                       &        & $H\parallel c$ 	&   4.14(3)	&  0.227(1) & 0.97(1)		&	      &    $-16.4(4)$ &		    \\
Ca(Co$_{0.48}$Ni$_{0.52}$)$_{1.87}$As$_2$      &     	& $H\parallel ab$ 	&  3.63(6) 	&  0.175(2) & 0.86(1)		& 0.89(1) &    $-45(2)$ &       	\\
						                       &        & $H\parallel c$ 	&   3.98(8)	&  0.211(3) & 0.94(1)		&	      &    $-46(2)$ &		    \\
Ca(Co$_{0.33}$Ni$_{0.67}$)$_{1.89}$As$_2$      &     	& $H\parallel ab$ 	&  3.73(3) 	&  0.128(1) & 0.73(1)		& 0.77(1) &    $-77(2)$ &       	\\
						                       &        & $H\parallel c$ 	&   3.66(8)	&  0.170(3) & 0.84(1)		&	      &    $-90(3)$ &		    \\
Ca(Co$_{0.19}$Ni$_{0.81}$)$_{1.94}$As$_2$      &     	& $H\parallel ab$ 	&  3.19(2) 	&  0.090(3) & 0.60(1)		& 0.65(1) &    $-203(16)$ &       	\\
						                       &        & $H\parallel c$ 	&   3.90(3)	&  0.140(2) & 0.75(1)		&	      &    $-128(3)$ &		    \\
\end{tabular}
\end{ruledtabular}
\end{table*}

\begin{figure}
\includegraphics[width =3.4in]{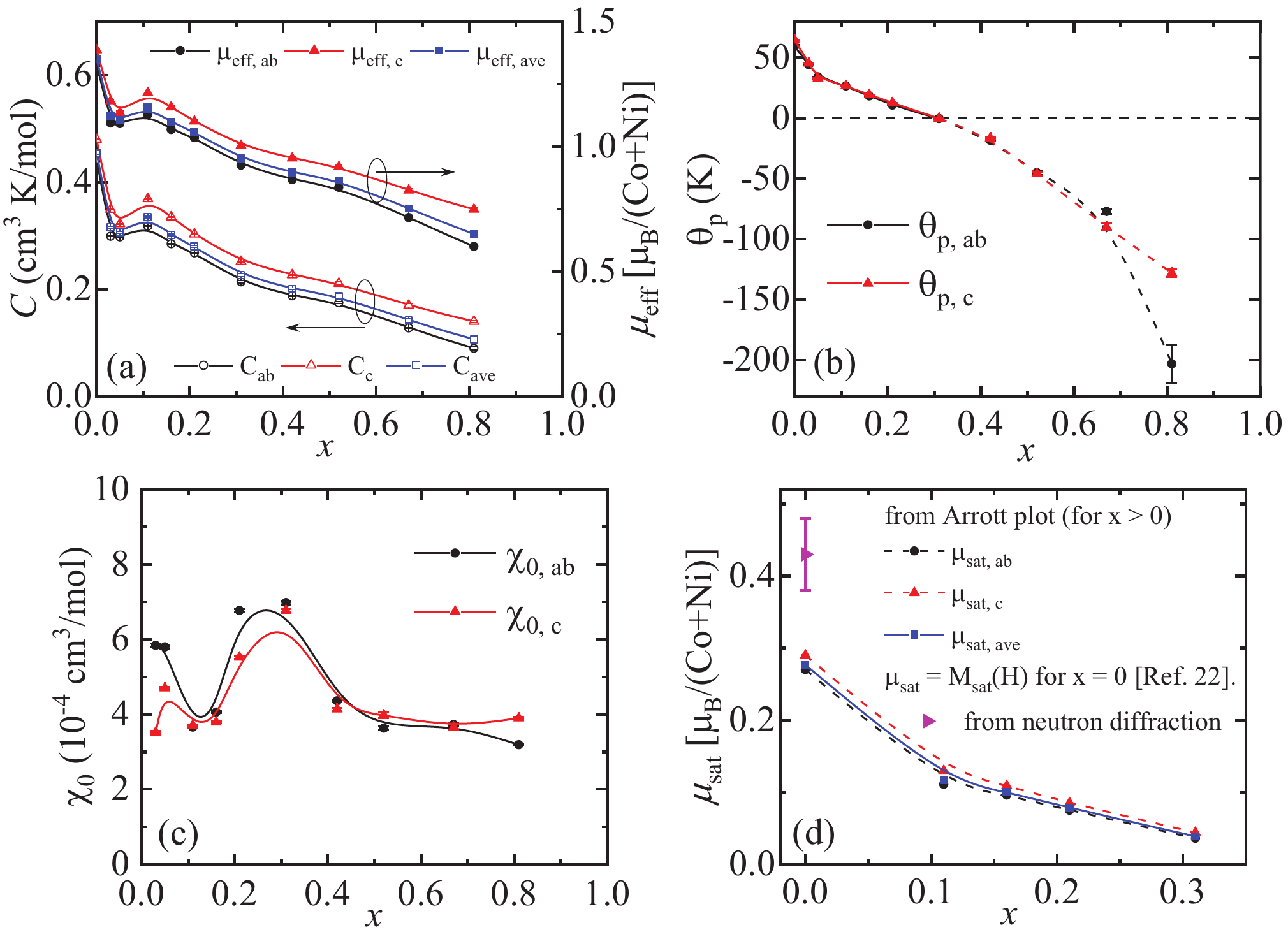}
\caption{Magnetic parameters versus~$x$ from Table~\ref{Tab.chidata}: (a)~Curie constants $C_{ab},\ C_c,\ C_{\rm ave}$ (left ordinate) and effective magnetic moments $\mu_{{\rm eff}\,{ab}},\ \mu_{{\rm eff}\,c},\ \mu_{\rm eff\,ave}$ (right ordinate), (b)~Weiss temperatures $\theta_{{\rm_p}\, ab},\ \theta_{{\rm p}\, c}$ and (c)~$T$-independent contributions $\chi_{0,\,ab},\ \chi_{0,\,c}$. The large negative values of $\theta_{\rm_{p}}$ for $x \geq 0.42$ are unphysical and are only to be considered as fitting parameters.  (d)~Saturation moments $\mu_{{\rm sat}\,ab}$, $\mu_{{\rm sat}\,c}$, and $\mu_{\rm sat\,ave}$ from Table~\ref{Tab.sfcoercive}.  The data for $x=0$ were obtained from zero-field neutron diffraction and $M$(2~K,~$H=14$~T) data.  The data for $x=0.11$--0.31 were obtained from the Arrott plots in the insets of Fig.~\ref{Fig_hysteresis_2K}.}
\label{Fig_Mag_parameters}
\end{figure}

\begin{figure}
\includegraphics[width = 3.in]{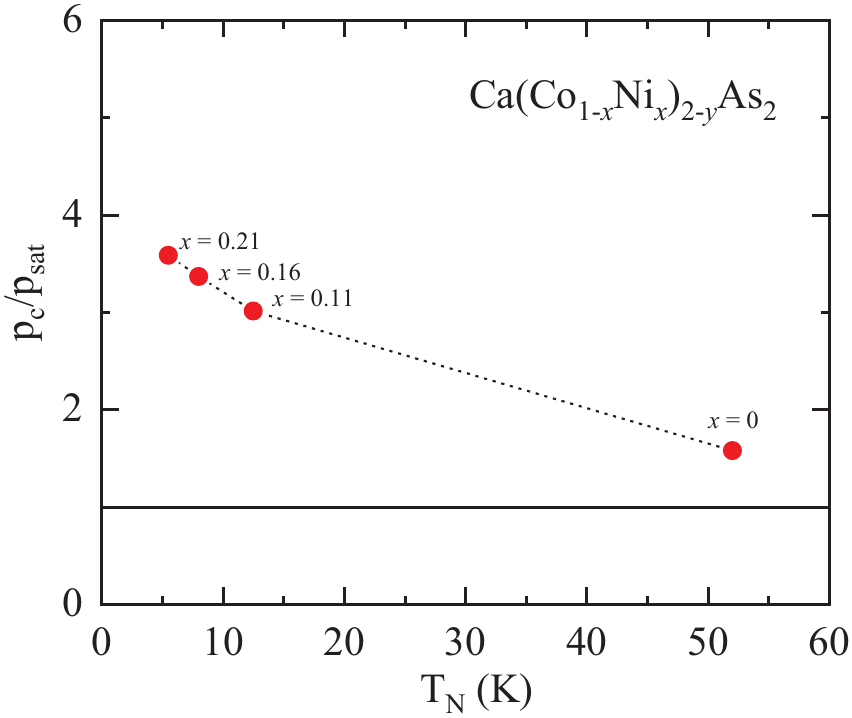}
\caption{Rhodes-Wohlfarth plot of $p_{\rm c}/p_{\rm sat}$ vs $T_{\rm N}$  for the listed \ccna\ compositions.  Although such plots are usually versus the FM Curie temperature $T_{\rm C}$ on the abscissa, here FM interactions dominate and hence the abscissa is the N\'eel temperature $T_{\rm N}$.  Because the crystals with $x=0.16$ and~0.21 do not show AFM ordering, proxies for the $T_{\rm N}$ values are used as explained in the text.}
\label{Fig_RW_plot}
\end{figure}

The $T$ dependences of the inverse magnetic susceptibilities $\chi_{ab}^{-1}(T)$ and $\chi_{c}^{-1}(T)$ for all the \ccna\ crystal compositions  $x = 0$--1 are shown in Figs.~\ref{Fig_Inversse_susceptibility}(a)--\ref{Fig_Inversse_susceptibility}(l). We analyzed the data in the PM regime $T>T_{\rm N}$ using the modified Curie-Weiss law for local magnetic moments given by
\bea
\chi_{\alpha}(T) =\chi_0 + \frac{C_{\alpha}}{T-\theta_{\rm p\alpha}} \qquad (\alpha = ab,\ c),
\label{Eq.ModCurieWeiss}
\eea
where $\chi_0$ is an isotropic $T$-independent term, $\theta_{\rm p\alpha}$ is the Weiss temperature for field direction~$\alpha$,   and the  Curie constant~$C_\alpha$ per mole of formula units (f.u.) is given by
\bea
C_{\alpha} = \frac{N_{\rm A} {g_\alpha}^2S(S+1)\mu^2_{\rm B}}{3k_{\rm B}} \quad (\alpha=ab,\ c),
\label{Eq.Cvalue1}
\eea
where $N_{\rm A}$ is Avogadro's number, $g_\alpha$ is the spectroscopic splitting factor ($g$ factor), $S$ is the spin angular-momentum quantum number, $k_{\rm B}$ is Boltzmann's constant, and $\mu_{\rm eff}$ is the effective moment of a spin.  The $\mu_{\rm eff}$  per (Co+Ni) transition metal atom is obtained from $C_\alpha$ per mole of formula units using
\bea
\mu_{\rm eff \alpha}{\rm [\mu_B/(Co+Ni)]} =  \sqrt{\frac{8C_\alpha{\rm [cm^3\,K/mol\,f.u.]}}{2-y}}.
\label{Eq.mueff}
\eea

As shown in Figs.~\ref{Fig_Inversse_susceptibility}(a)--\ref{Fig_Inversse_susceptibility}(k), the $\chi_{ab}^{-1}(T)$ and $\chi_{c}^{-1}(T)$ data are well described by Eq.~(\ref{Eq.ModCurieWeiss}) in the $T$ range noted in the panel for each crystal and the fitted parameters are listed in Table~\ref{Tab.chidata}. The values of $C_{\alpha}$, $\mu_{\rm eff\,\alpha}$, $\theta_{\rm_{p\alpha}}$, and $\chi_{0\alpha}$ as a function of~$x$ for the \ccna\ crystals are plotted in \mbox{Figs.~\ref{Fig_Mag_parameters}(a)--\ref{Fig_Mag_parameters}(c)}. The $C_{\alpha}$ and $\mu_{\rm eff\alpha}$ decrease with increasing $x$.  The positive $\theta_{\rm_{p\alpha}}$ also decreases with increasing $x$ and becomes negative for $x \geq 0.31$. The large negative value of $\theta_{\rm_{p\alpha}}$ for $x \geq 0.41$ is a result of the overall suppression of $\chi(T)$ at the larger $x$ values and does not have physical significance.

In a local-moment picture the isotropic Curie constant in units of ${\rm cm^3\,K/mol\,spins}$ for Heisenberg spins~$S$ with $g=2$ is given by
\bea
C\left({\rm \frac{cm^3\,K}{mol\,spins}}\right) = 0.5002 S(S+1).
\label{Eq.Curieconstant}
\eea
Thus for the minimum $S = 1/2$ with \mbox{$g=2$} and two spins per formula unit, one obtains \mbox{$C_{\rm mol} = 0.75~{\rm cm^3\,K/(mol\,f.u.})$}. The Curie constants for all the \ccna\ crystals in Table~\ref{Tab.chidata} are much smaller than this value and hence suggest an itinerant character of the magnetism in these crystals as in undoped \cca\ where FM interactions dominate.  For an itinerant weak ferromagnet, the degree of itinerancy can be expressed in a Rhodes-Wohlfarth (RW) plot~\cite{Rhodes1963,Santigo2017} of the ratio $p_{\rm c}/p_{\rm sat}$ versus the FM Curie temperature $T_{\rm C}$, where  $p_{\rm sat} = \mu_{\rm sat}/\mu_{\rm B}$ is the saturation moment per transition-metal atom normalized by the Bohr magneton and~\cite{Sangeetha2019scna}
\bea
p_{\rm c} = \sqrt{1 + p_{\rm eff}^2} - 1 \qquad (g=2).
\label{Eq:p_c}
\eea

For Heisenberg local-moment spins with $g=2$ the value of $p_{\rm c}/p_{\rm sat}$ is equal to unity, whereas a larger value of $p_{\rm c}/p_{\rm sat}$ signifies itinerant magnetic character.  Figure~\ref{Fig_RW_plot} shows an RW plot of $p_{\rm c}/p_{\rm sat}$ versus $T_{\rm N}$ for four of our \ccna\ crystals, where the $p_{\rm c}$ values are listed in Table~\ref{Tab.chidata}.  The value of $p_{\rm sat}$ is estimated by plotting $M^2$ versus $H/M$ at $T\ll T_{\rm C}$ (i.e., $\ll T_{\rm N}$ here) known as an Arrott plot~\cite{Arrott1957, Arrott2010} and extrapolating the high-field data to $H/M=0$~\cite{Takahashi2013} as shown in the insets in Fig.~\ref{Fig_hysteresis_2K} below. For \cca\ $(x=0)$ the 
saturation (ordered) moment is taken to be that obtained from zero-field neutron diffraction measurements~\cite{Jayasekara2017}. We have seen that AFM ordering does not occur for $x \geq 0.16$.  Therefore proxies for the values of $T_{\rm N}$ for $x = 0.16$ and 0.21 in the RW plot are taken as the temperatures where the curvature of $\chi_c(T)$ in Figs.~\ref{Fig_M-T_all_separate}(e) and~\ref{Fig_M-T_all_separate}(f) respectively change sign from positive to negative on cooling. As seen from Fig.~\ref{Fig_RW_plot}, $p_{\rm c}/p_{\rm sat} > 1$ for all measured \ccna\ compositions and also $p_{\rm c}/p_{\rm sat}$ increases with increasing $x$. This behavior suggests that the \ccna\ crystals are essentially weak itinerant ferromagnets as expected from the dominance of FM interactions in these materials as also previously inferred for the \scna\ system~\cite{Sangeetha2019scna}.

\subsection{Magnetization versus applied magnetic field isotherms}

\begin{figure*}
\includegraphics[width = \textwidth]{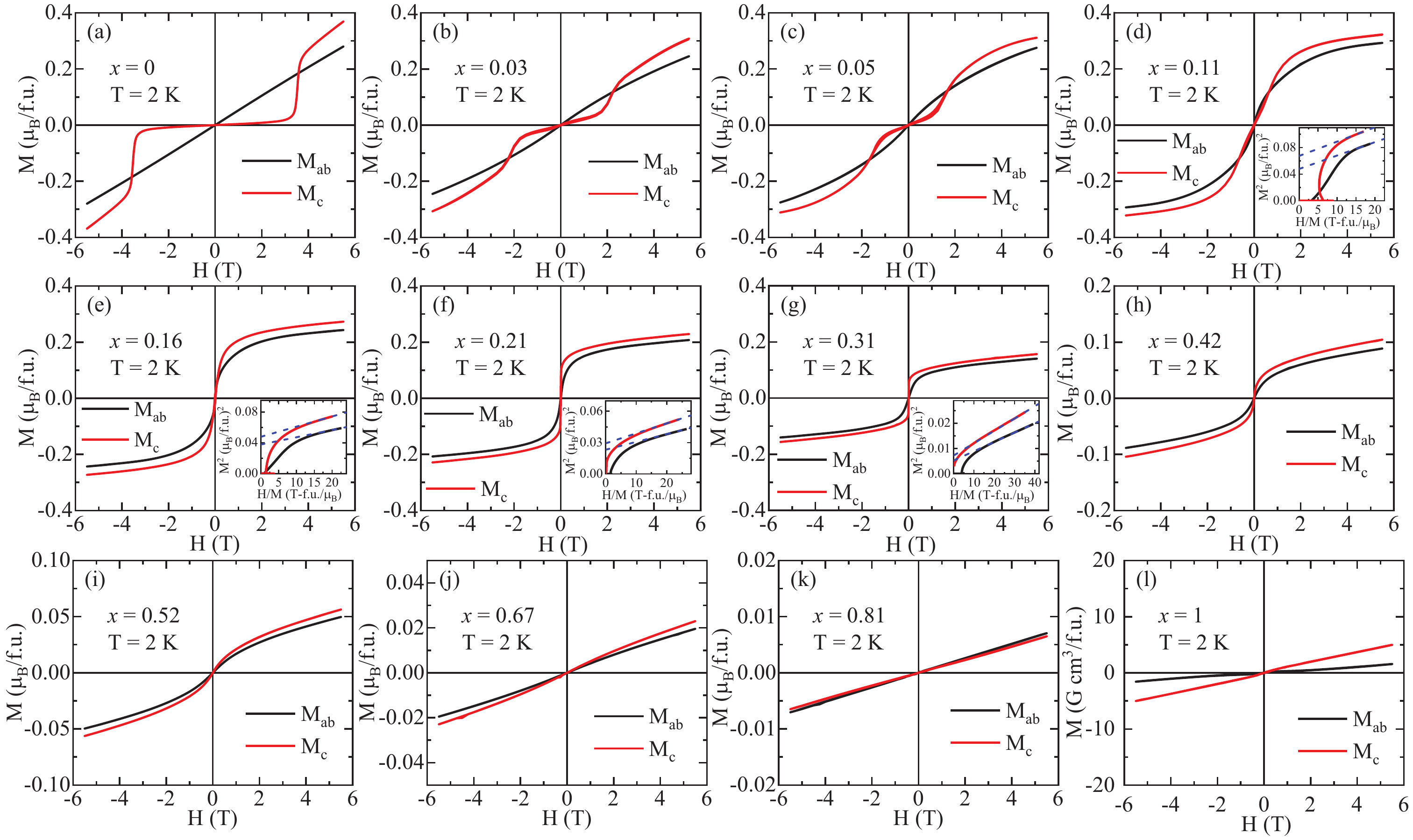}
\caption{Four-quadrant magnetic hysteresis measurements at $T=2$~K of \ccna\ crystals with $x = 0$ to 1 as indicated,  measured for $H \parallel ab$ ($M_{ab}$) and $H \parallel c$ ($M_c$).  Insets~(d)--(g): Arrott plots of $M^2$ vs $H/M$ for both $H \parallel ab$ and $H \parallel c$.  The linear highest-field linear $M^2(H/M)$ data in each panel are extrapolated to $H=0$ to obtain the value of $M(H=0,T=0)$ and then the saturation (ordered) moment per transition-metal atom $\mu_{\rm sat} = M(H=0,T=0)/(2-y)$.}
\label{Fig_hysteresis_2K}
\end{figure*}

\begin{figure}
\includegraphics[width = 2.5in]{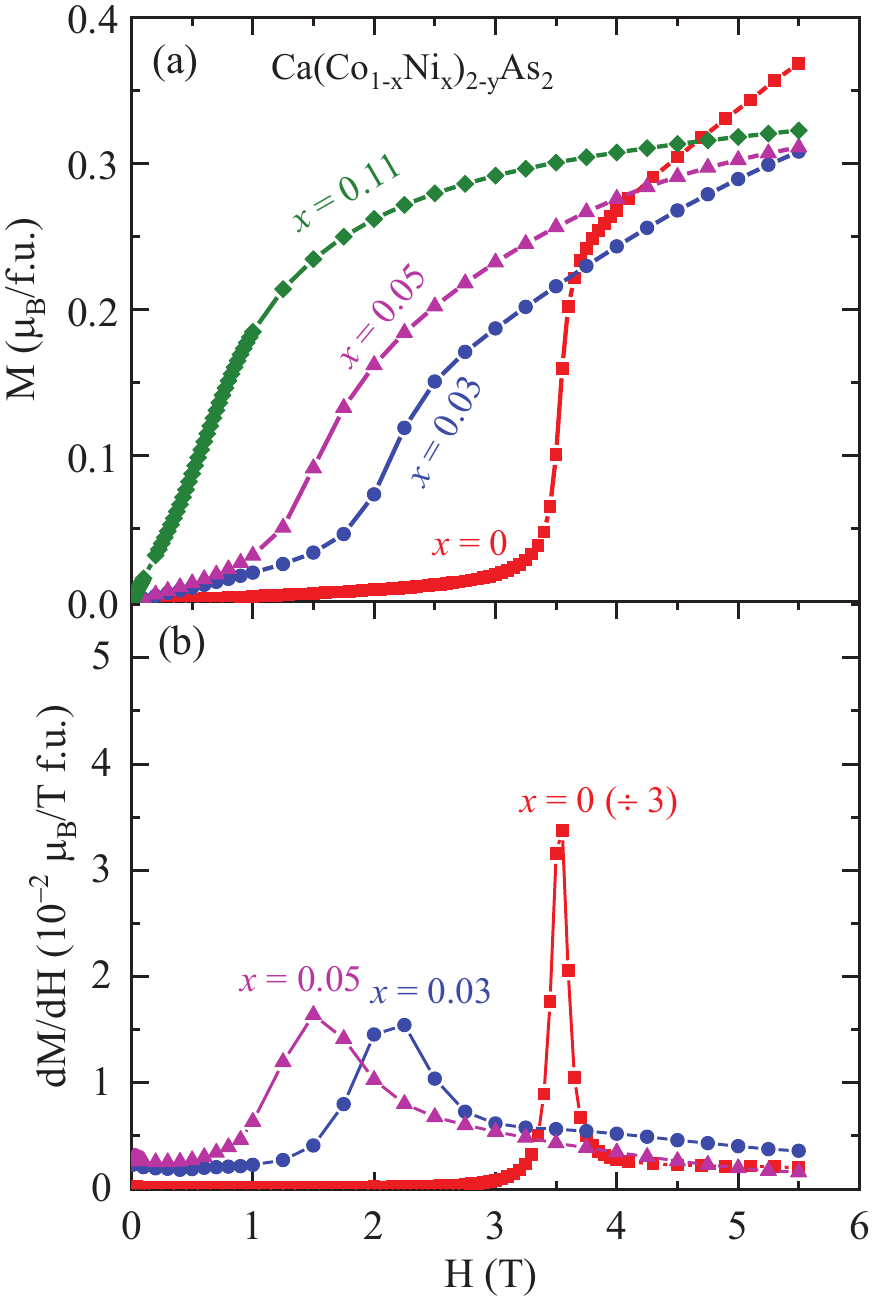}
\caption{(a)~Expanded plots of the magnetization $M$ vs applied magnetic field \mbox{$H\parallel c$} for \ccna\ crystals with x = 0, 0.03, 0.05, and 0.11 from Figs.~\ref{Fig_hysteresis_2K}(a)--\ref{Fig_hysteresis_2K}(d), respectively.  (b)~Derivative $dM/dH$ of the $M(H)$ data in~(a).  Note that the peak height for $x=0$ is divided by 3.}
\label{Fig_M-H_0-3-5-11}
\end{figure}

In order to investigate the magnetic-field evolution of the magnetic ground state, magnetization versus applied magnetic field $M(H)$ hysteresis measurements were carried out for each of our crystal compositions.  Before each measurement the superconducting magnet was quenched to reduce the applied field to zero.  \mbox{Figures}~\ref{Fig_hysteresis_2K}(a)--\ref{Fig_hysteresis_2K}(l) show $M(H)$ four-quadrant hysteresis data with $H \parallel ab$ and $H \parallel c$ measured at $T = 2$~K for each of the \ccna\ crystals. 

As reported earlier~\cite{Anand2014Ca, Cheng2012b, Ying2012}, the undoped $x = 0$ composition shows a clear spin-flop (SF) transition in the $M_c(H)$ data at  $H_{\rm SF} = 3.6$~T, consistent with the known collinear \mbox{$c$-axis} AFM ordering~\cite{Jayasekara2017}, whereas $M_{ab}(H)$ shows a smooth behavior. Figure~\ref{Fig_M-H_0-3-5-11}(a) shows expanded plots of the $M_c(H)$ data in Fig.~\ref{Fig_hysteresis_2K} for $x = 0$, 0.03, 0.05, and 0.11.  The latter three compositions show spin-flop transitions.  The derivatives $dM/dH$ versus~$H$ for these three compositions are plotted in Fig.~\ref{Fig_M-H_0-3-5-11}(b) from which we obtain the respective spin-flop fields $H_{\rm SF}$  listed in Table~\ref{Tab.sfcoercive}.

\begin{table}
\caption{\label{Tab.sfcoercive} Spin-flop transition field  $H_{\rm SF}$, coercive field $H_{\rm CF}$, remanent magnetization $M_{\rm rem}$, FM ordered moment $\mu_{\rm FM}$, and $p_{\rm sat}$ values of the \ccna\ crystals with $x=0$--0.31.  The values of $p_{\rm sat}\equiv\mu_{\rm sat}/\mu_{\rm B}$ in units of $\mu_{\rm B}$ per transition metal (TM) atom were obtained from the $M^2$ vs $H/M$ isotherms in the insets of Figs.~\ref{Fig_hysteresis_2K}(d)--\ref{Fig_hysteresis_2K}(g). The $p_{\rm sat}$ values for $x=0$ were obtained from $M$(2~K,$H=14$~T) data~\cite{Anand2014} and $H=0$, $T = 4$~K neutron-diffraction data~\cite{Jayasekara2017} and the values for $x>0$ from Arrott plots.  The $H_{\rm CF}$, $M_{\rm rem}$ and $\mu_{\rm FM}$ values associated with FM ordering were obtained from the data in Fig.~\ref{Fig_hysteresis_11-16-21-31}. }
\begin{ruledtabular}
\begin{tabular}{ccccccc}	
$x$	& $H$		& $H_{\rm SF}$	 &  $H_{\rm CF}$& $M_{\rm rem}$	& $\mu_{\rm FM}$	&$p_{\rm sat}$ \\
	& direction	& (T)			& (Oe)		& ($\mu_{\rm B}$/f.u.)& ($\mu_{\rm B}$/f.u.)&  ($\mu_{\rm B}$/TM) \\
\hline
0	& $H =0$			& 		&   			&  			&				& 0.43(5) \cite{Jayasekara2017}	\\
	& $H \parallel ab$	& 		&   			&  			&				& 0.27 \cite{Anand2014}			\\
    	& $H \parallel c$	& 3.60(1) 	&   			& 			&   				& 0.29 \cite{Anand2014}			\\
0.03	& $H \parallel ab$	& 		&   			&   			&    				&				\\
    	& $H \parallel c$	& 2.2(1) 	&   			&   			&    				&				\\
0.05	& $H \parallel ab$	& 	 	&   			&   			&    				&				\\
    	& $H \parallel c$	& 1.5(1) 	&   			&   			&    				&				\\
0.11	& $H \parallel ab$	& 	 	&  $\sim$ 0 	&   	0		&	 $\sim$ 0		&   0.176(2)		\\
    	& $H \parallel c$	&	 	&  150(5) 		&   	0.0032	&	0.01			&   				\\
0.16	& $H \parallel ab$	& 	 	&  20(5) 		&   	0.0025	&	0.043		&   0.144(2)		\\
    	& $H \parallel c$	&  		&  160(5) 		&   	0.0139	&	0.03			&   				\\
0.21	& $H \parallel ab$	& 	 	&  $\sim$ 0 	&   	0.0020	&	0.034		&   0.127(2)		\\
    	& $H \parallel c$	&  		&  25(5) 		&   	0.0163	&	0.11			&   				\\
0.31	& $H \parallel ab$	& 	 	&  $\sim$ 0 	&   	0		&	0.009		&   0.094(2)		\\
    	& $H \parallel c$	&  		&  $\sim$ 0	&   	0.0103	&	0.067		&   				\\
\end{tabular}
\end{ruledtabular}
\end{table}

\begin{figure}
\includegraphics[width = 2.5in]{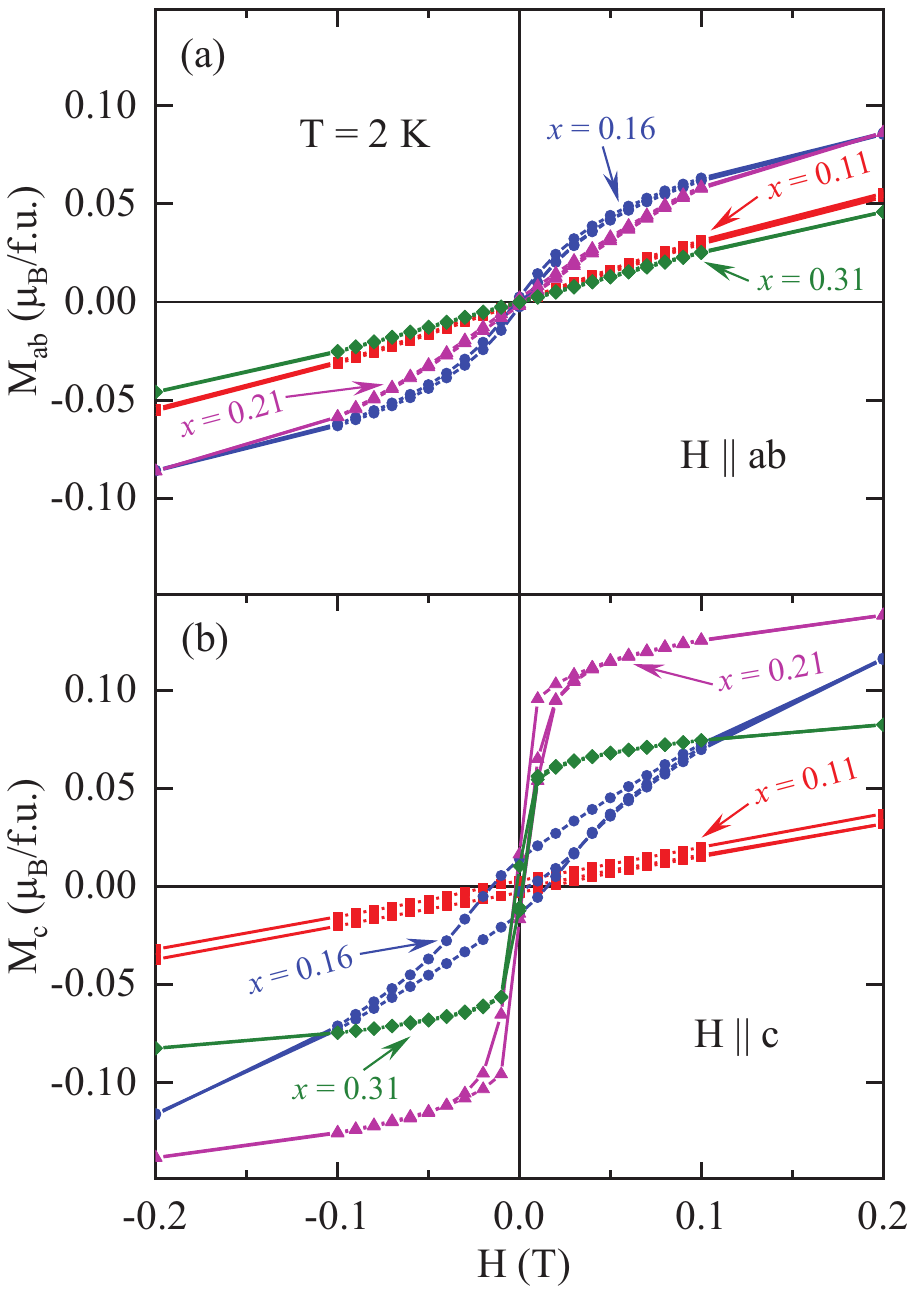}
\caption{Expanded low-field plots at $T=2$~K of the $M(H)$ hysteresis loops of \ccna\ crystals with $x$ = 0.11, 0.16, 0.21, and 0.31 for (a)~$H \parallel ab$ and (b)~$H \parallel c$.}
\label{Fig_hysteresis_11-16-21-31}
\end{figure}

No magnetic hysteresis is observed in Fig.~\ref{Fig_hysteresis_2K} in either $M_{ab}(H)$ or $M_c(H)$ for $x = 0$, 0.03, 0.05 or $x \geq 0.41$.  However hysteresis in the $M_c(H)$ isotherms at low fields was detected at $T=2$~K for the crystals with $0.11 \leq x \leq 0.31$, as shown in Figs.~\ref{Fig_hysteresis_11-16-21-31}(a) and~\ref{Fig_hysteresis_11-16-21-31}(b) for $H\parallel ab$ and $H\parallel c$, respectively.  These data indicate the occurrence of weak ferromagnetism and the parameters associated with it are the coercive field $H_{\rm CM}$, remanent magnetization $M_{\rm rem}$ and FM moment $\mu_{\rm FM}$ listed in Table~\ref{Tab.sfcoercive}.  The observation of small $H_{\rm CF}$ values for $x = 0.11$ and 0.16 and even smaller values for $x=0.21$ and 0.31 is consistent with the notion that the suppression of AFM ordering is mediated by strong FM fluctuations. The quasi-1D $c$-axis FM fluctuations discussed earlier are also reflected in the $M(H)$ behavior, where only the $M_c(H)$ data are hysteretic to some extent, whereas hysteresis is much smaller in the $M_{ab}(H)$ data.

We obtained estimates of the saturation moments for $x=0.11$, 0.16, 0.31, and 0.41 from extrapolations to $H=0$ of Arrott plots of $M^2$ versus $H/M$ shown in the respective insets of Fig.~\ref{Fig_hysteresis_2K}.  The results are listed in Table~\ref{Tab.sfcoercive} and plotted in Fig.~\ref{Fig_Mag_parameters}(d) together with the \mbox{low-$T$} ordered-moment for $x=0$ obtained from zero-field neutron-diffraction~\cite{Jayasekara2017} and $M(H=14~$T,~$T=2$~K)~\cite{Anand2014Ca} measurements.

\section{\label{Sec:Cp} Heat capacity}

\begin{figure*}
\includegraphics[width = \textwidth]{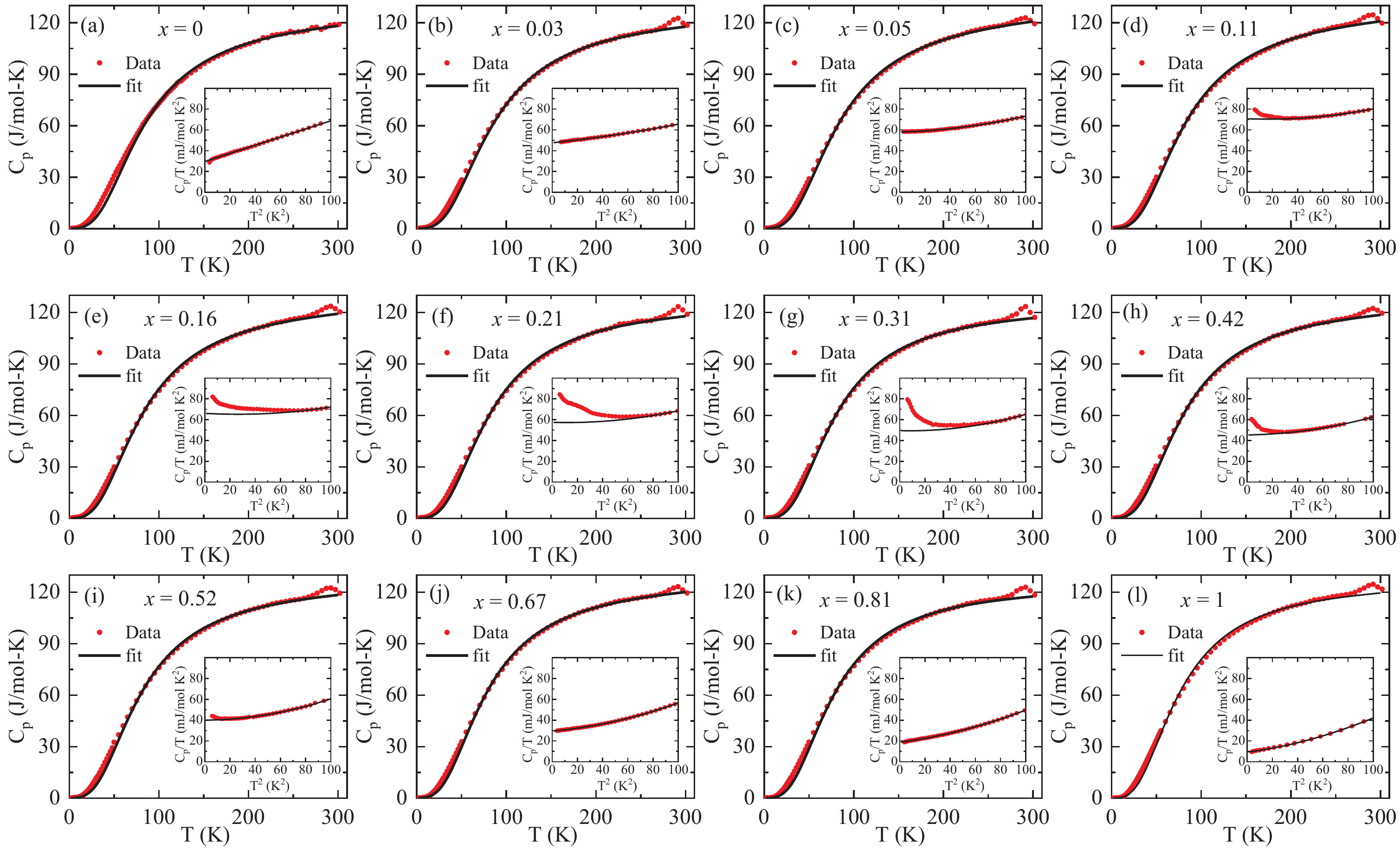}
\caption{Temperature dependence of zero-field heat capacity $C_{\rm p}(T)$ of \ccna\ crystals with $x = 0$--1 along with fits by the Debye model using Eq.~(\ref{Eq:Debye_Fit}).  The humps in $C_{\rm p}(T)$ at $T \approx 290$~K are experimental artifacts.  Insets: $C_{\rm p}(T)/T$ vs $T^2$ along with fits by Eq.~(\ref{Eq.CpFit1}). The fitting temperature range is 2--10~K for $x = 0$, 0.03, 0.05, 0.67, 0.81, and 1, whereas for all other compositions the range is $T_{\rm min} \leq T \leq 10$~K, with $T_{\rm min}$ being the temperature where minima in $C_{\rm p}(T)/T$ vs $T^2$ is observed and below which the low-$T$ upturn occurs that is associated with FM fluctuations. The fits below $T_{\rm min}$ are the extrapolations of the fits to 2 K using Eq.~(\ref{Eq.CpFit1}).}
\label{Fig_Cp_T_All}
\end{figure*}

\begin{table*}
\caption{\label{Tab.heatcapacityfit} The fitting parameters obtained from the analysis of heat capacity data. Listed are the crystal composition, doping charge $q$ with respect to ${\rm CaCo_2As_2}$ from Eq.~(\ref{Eq:DopingLevel}) (hole-doping values are negative),  Sommerfeld coefficient $\gamma$ obtained using Eq.~(\ref{Eq.CpFit1}) for $x = 0$, 0.03, 0.05, 0.67, 0.81, and 1 and using Eq.~(\ref{Eq.Cp_SF_Fit}) for $x = 0.11$, 0.16, 0.21, 0.31, 0.41, and 0.52; lattice heat-capacity coefficients $\beta$ and $\delta$ obtained from low-$T$ fits of the C$_{\rm p}$/$T$ versus $T^2$ data; spin-fluctuation coefficient $\kappa$, spin fluctuation temperature $T_{\rm sf}$, and density of states at the Fermi energy ${\cal D}(E_{\rm F})$ derived from $\gamma$ using Eq.~(\ref{Eq:DfromGamma}).}
\begin{ruledtabular}
\begin{tabular}{cccccccc}	
  	Compound		& doping~$q$ & $\gamma$	& $\beta$			& $\delta$ 	& $\kappa$ &  $T_{\rm {sf}}$ & ${\cal D}(E_{\rm F})$ \\
 					&	&	(mJ/mol\,K$^{2}$) & (mJ/mol\,K$^{4}$)	& ($\mu$J/mol\,K$^{6}$)	 & (mJ/mol\,K$^{2}$)  & (K) & [states/(eV\,f.u.)] \\
\hline

CaCo$_{1.86(2)}$As$_2$     & $-0.98$                    & 	29.6(1) 	& 0.391(1)	& $\approx 0$ 	&  	&  		& 12.54(5)  \\

Ca(Co$_{0.97}$Ni$_{0.03}$)$_{1.86}$As$_2$   & $-0.92$    & 	47.59(8) 	& 0.126(4)	&   0.52(4)	&   	&  		& 20.17(3) \\

Ca(Co$_{0.95}$Ni$_{0.05}$)$_{1.86}$As$_2$    & $-0.89$   & 	57.98(1) 	& 0.131(3) 	&   1.21(3)	&   	&  		& 24.57(1) \\

Ca(Co$_{0.89}$Ni$_{0.11}$)$_{1.87}$As$_2$     & $-0.70$  &    53.5(7) & 0.21(1)	&   0.44(7)	&   $-16.9(5)$	&  	10.4(8)	& 22.6(3) 	\\

Ca(Co$_{0.84}$Ni$_{0.16}$)$_{1.89}$As$_2$      & $-0.47$ &    59(1) 	& 0.14(4) 	&   0.8(2) 	&  $-15(1)$  &  10.21(3)		&  25.0(4)	\\

Ca(Co$_{0.79}$Ni$_{0.21}$)$_{1.86}$As$_2$      & $-0.59$ &    55(1)  	& 0.14(3) 	&   1.3(1) 	&  $-24(2)$  &  8.14(2)		&  23.3(4)	\\

Ca(Co$_{0.69}$Ni$_{0.31}$)$_{1.87}$As$_2$      & $-0.33$&    52(1) 	& 0.24(3) 	&   0.6(1) 	&  $-49.2(7)$  &  7(1)		& 22.0(4)	\\

Ca(Co$_{0.58}$Ni$_{0.42}$)$_{1.87}$As$_2$     & $-0.12$  &    40(1) 	& 0.35(1) 	&   0.37(6) 	& $-21.6(4)$   &  4.80(6)		&  16.9(4) 	\\

Ca(Co$_{0.48}$Ni$_{0.52}$)$_{1.87}$As$_2$     & $0.06$  &    32.9(4) 	& 0.22(1)	&  0.90(5)  	& $-9.8(4)$   &  	4(1)	& 13.9(2) 	\\

Ca(Co$_{0.33}$Ni$_{0.67}$)$_{1.89}$As$_2$    & $0.50$   &    29.62(4)  	& 0.10(3)	&   1.67(2) 	&    &  		&  12.55(2)   \\

Ca(Co$_{0.19}$Ni$_{0.81}$)$_{1.94}$As$_2$    & $1.15$   &    18.77(6) 	& 0.15(3) 	&   1.51(3) 	&    &  		&  7.95(3) 	\\

CaNi$_{1.97}$As$_2$       & $1.76$                      & 	9.3(1) 	& 0.15(6) 	& 1.7(6) 	&   	&  	&  3.94(4)   \\

\end{tabular}
\end{ruledtabular}
\end{table*}

\begin{figure}
\includegraphics[width = 0.48\textwidth]{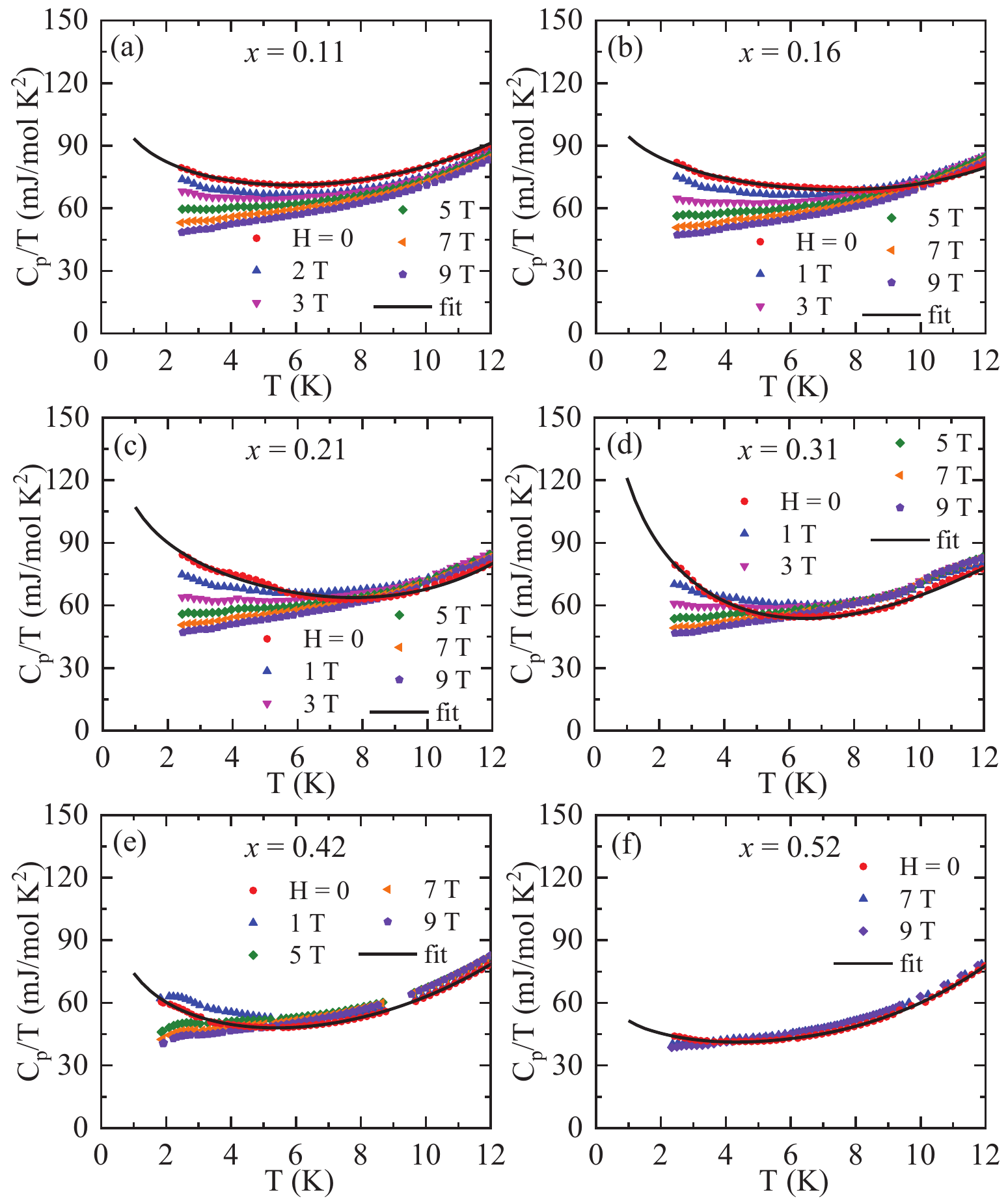}
\caption{$C_{\rm p}/T$ vs $T$ data of \ccna\ crystals with $x$ = 0.11 (a), 0.16 (b), 0.21 (c), 0.31 (d), 0.42 (e), and 0.52 (f), measured at different magnetic fields applied along the $c$~axis. The zero-field data are fitted in the temperature range 2--12~K by Eq.~(\ref{Eq.Cp_SF_Fit}) as shown by the black solid lines. The \mbox{low-$T$} upturns are associated with FM quantum spin fluctuations that are suppressed with increasing magnetic field.}
\label{Fig_Cp_quantum_critical_fit_with_field}
\end{figure}

\begin{figure}
\includegraphics[width = 0.48\textwidth]{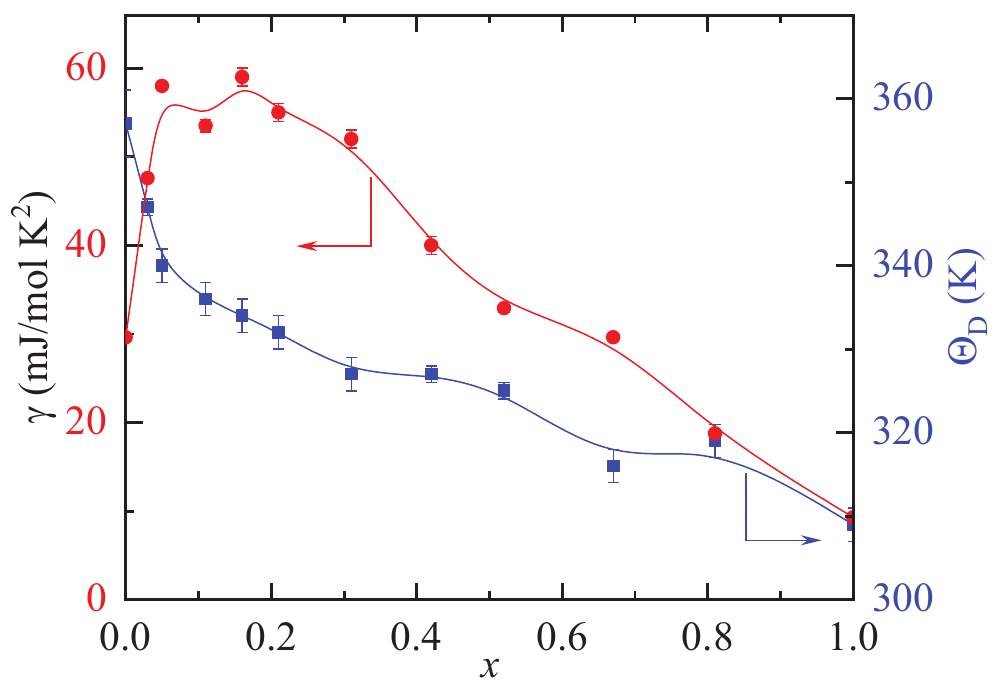}
\caption{Sommerfeld coefficients $\gamma$ from Table~\ref{Tab.heatcapacityfit}  of \ccna\ crystals with $x = 0$, 0.03, 0.05, 0.67, 0.81, and 1 from low-$T$ heat capacity fits using Eq.~(\ref{Eq.CpFit1}) and for $x = 0.11$, 0.16, 0.21, 0.31, 0.41, and 0.52 using Eq.~(\ref{Eq.Cp_SF_Fit}) (filled red circles). The Debye temperatures $\Theta_{\rm D}$ for all compositions from Table~\ref{Tab.heatcapacityDebyefit} were determined by fitting the $C_{\rm p}(T)$ data by Eq.~(\ref{Eq:Debye_Fit}) in the $T$ range 50--300~K (filled blue squares). The lines are guides to the eye.}
\label{Fig_Cp_parameters}
\end{figure}

The zero-field heat capacities $C_{\rm p}(T)$ for all the \ccna\ crystal compositions are shown in Figs.~\ref{Fig_Cp_T_All}(a)--\ref{Fig_Cp_T_All}(l) in the $T$ range 2--300~K\@. No detectable feature is observed at $T_{\rm N}$ for any of the respective compositions, evidently due to the small magnetic entropy change in these itinerant magnets at $T_{\rm N}$\@. The $C_{\rm p}(T)$ value at our high-$T$ limit of 300~K is close to the classical Dulong-Petit high-$T$ limit $C_{\rm p} = 3nR = 124.7$~J/mol\,K where $R$ is the molar gas constant and here $n \approx 5$ is the number of atoms per formula unit.  As seen from the figures for all compositions except $x = 0$, a hump-like feature appears in the $C_{\rm p}(T)$ at $T \sim 290$~K that is due to melting of the Apiezon~N grease used to make thermal contact between the sample and platform of the sample puck. This spurious feature can be avoided by using Apiezon H grease for the high-$T$ measurements as shown for the crystal with $x = 0$.

Due to the nonstoichiometry of the \ccna\ crystals, the charge doping with respect to stoichiometric ${\rm CaCo_2As_2}$ is not equal to the value of $x$.  Instead, noting that Co$^{2+}$ and Ni$^{2+}$ respectively have $3d^7$ and $3d^8$ electron configurations, the doping charge per formula unit $q$ in units of the elementary charge is given by
\bea
q = [7(1-x) +8x](2-y)-14,
\label{Eq:DopingLevel}
\eea
where a negative value of $q$ corresponds to hole doping and a positive value to electron doping.  The doping levels per formula unit are given for each of the crystal compositions in the second column of Table~\ref{Tab.heatcapacityfit}.

The insets of Figs.~\ref{Fig_Cp_T_All}(a)--\ref{Fig_Cp_T_All}(l) show $C_{\rm p}(T)/T$ vs $T^2$ for all the \ccna\ crystal compositions. The low-$T$ data in the range 2--10~K for $x = 0$, 0.03, 0.05, 0.67, 0.81, and 1 were fitted by the expression
\bea
C_{\rm p}(T) = \gamma T+ \beta T^3 +\delta T^5,
\label{Eq.CpFit1}
\eea
where $\gamma$ is the Sommerfeld electronic heat-capacity coefficient associated with the conduction carriers and the last two terms describe the low-$T$ lattice heat-capacity contribution. The fitted parameters are listed in Table~\ref{Tab.heatcapacityfit}.  As shown in Fig.~\ref{Fig_Cp_T_All}, the $C_{\rm p}(T)$ vs $T^2$ data for the $x = 0.11$, 0.16, 0.21, 0.31, 0.42, and 0.52 crystals exhibit an upturn below a temperature $T_{\rm min}(x)$ below which the $C_{\rm p}(T)$ vs $T^2$ data are not described by Eq.~(\ref{Eq.CpFit1}).

As discussed earlier, the AFM interactions are gradually suppressed with increasing Ni substitution in the \ccna\ crystals. For the $x = 0.11$ and 0.16 crystals the strengths of the AFM and FM interactions are comparable so that the materials settle into a metastable state below their respective $T_{\rm N}$\@. The FM spin fluctuations dominate the AFM fluctuations with a further increase in the Ni concentration as inferred from a strong signature of FM spin fluctuations observed in the susceptibility data for $x = 0.21$ and 0.31 giving rise to a low-$T$ upturn in the $C_{\rm p}/T$ vs $T^2$ data as shown in the insets of Fig.~\ref{Fig_Cp_T_All} and further discussed below.

To elucidate and confirm the presence of FM spin fluctuations in the  $x = 0.11$, 0.16, 0.21, 0.31, 0.41, and 0.52 crystals, we measured $C_{\rm p}/T$ versus $T$ for magnetic fields applied along the $c$~axis and the corresponding plots are shown in Figs.~\ref{Fig_Cp_quantum_critical_fit_with_field}(a)--\ref{Fig_Cp_quantum_critical_fit_with_field}(f). One may argue that such upturns may also be associated with the high-$T$ tail of a Schottky anomaly; however, in this case an upturn is expected to increase with increasing magnetic field which conflicts with the observed data. In particular, the upturns in the $C_{\rm p}(T)$ data are gradually suppressed with increasing magnetic field.

In a magnetic system with FM quantum fluctuations a $C/T \sim \ln T$ contribution is observed~\cite{Sangeetha2019scna, Santanu2020CCIA, Flude1968, Millis1993, Nicklas1999, Maple2010, WuPNAS2014, Pandey2018}. We found that the zero-field $C_{\rm p}(T)$ at low~$T$ of the $x = 0.11$, 0.16, 0.21, 0.31, 0.41, and 0.52 crystals can also be well described by
\bea
\frac{C_{\rm p}(T)}{T} = \gamma + \beta T^2 +\delta T^4 + \kappa{\rm ln}(T/T_{\rm sf}),
\label{Eq.Cp_SF_Fit}
\eea
where in addition to the Sommerfeld coefficient $\gamma$ and lattice coefficients $\beta$ and $\delta$, $\kappa$ is the spin fluctuation coefficient and $T_{\rm sf}$ is the spin-fluctuation temperature. The fitted parameters are listed in Table~\ref{Tab.heatcapacityfit}. The FM spin fluctuations are observed over a wide composition range in \ccna\ in contrast to the ranges observed in isostructural \scna~\cite{Sangeetha2019scna} and \ccia~\cite{Santanu2020CCIA} crystals where the spin fluctuations were limited to narrow composition ranges.

The Sommerfeld coefficient $\gamma$ in Table~\ref{Tab.heatcapacityfit} increases by nearly a factor of two with only 3--5\% Ni substitutions, indicating a corresponding rapid increase in the electronic density of state at the Fermi energy ${\cal D}(E_{\rm F})$\@. The enhanced $\gamma$ values are observed for the compositions with strong FM fluctuations.  As the strength of these fluctuations gradually decreases for the larger $x$ values, $\gamma$ correspondingly decreases as shown in Fig.~\ref{Fig_Cp_parameters}. The ${\cal D}(E_{\rm F})$ values are determined from the $\gamma$ values using the relation
\bea
{\cal D}_\gamma(E_{\rm F})\rm  {\left(  \frac{states}{eV\,f.u.} \right) = \frac{1}{2.357}\ \gamma \left(\frac{mJ}{mol\,K^2}\right)   },
\label{Eq:DfromGamma}
\eea
and are listed in Table~\ref{Tab.heatcapacityfit}. Here, ${\cal D}_\gamma(E_{\rm F})$ is the density of states at $E_{\rm F}$ determined from $C_{\rm p}(T)$ measurements and includes the factor of two Zeeman degeneracy of the conduction carriers.

The $C_{\rm p}(T)$ data in the temperature range 50--300~K are analyzed according to
\bea
C_{\rm p}(T) &=& \gamma_{\rm D} T+ nC_{\rm V\,Debye}(T),\label{Eq:Debye_Fit} \\*
C_{\rm V\,Debye}(T) &=& 9R \left(\frac{T}{\Theta_{\rm D}}\right)^3\int_{0}^{\Theta_{\rm D}/T}\frac{x^4e^x}{(e^x-1)^2} dx,\nonumber
\eea
where $\gamma_{\rm D}$ is the Sommerfeld coefficient which we use as a variable in this fit, $n$ is the number of atoms per formula unit, $C_{\rm V\,Debye}$ is the Debye lattice heat capacity per mole of atoms at constant volume, and $\Theta_{\rm D}$ is the Debye temperature. The high-$T$ data associated with the phonon contribution are well described by the Debye model as shown by the black solid lines in Figs.~\ref{Fig_Cp_T_All}(a)--\ref{Fig_Cp_T_All}(l)\@. The fitted parameters are listed in Table~\ref{Tab.heatcapacityDebyefit}. From the Table, $\gamma_{\rm D}$ significantly differs from $\gamma$ in Table~\ref{Tab.heatcapacityfit} except for $x=0$ and $x=1$.

\begin{table}[h]
\caption{\label{Tab.heatcapacityDebyefit} Sommerfeld electronic heat-capacity coefficient $\gamma_{\rm D}$ and Debye temperature $\Theta_{\rm D}$ determined by fitting the $C_{\rm p}(T)$ data for \ccna\ crystals in the temperature range \mbox{50--300~K} by Eqs.~(\ref{Eq:Debye_Fit}). }
\begin{ruledtabular}
\begin{tabular}{ccc}	
  	$x$	 &  $\gamma_{\rm D} \left(\rm{mJ/mol\, K^2}\right)$ 	& $\Theta_{\rm D}$ (K) \\
\hline
0                      & 	29.4(3)		& 	357(4)	\\
0.03                   & 	28(1)		& 	347(1)	\\
0.05                   & 	22(1)		& 	340(2)	\\
0.11                   & 	22(1)		& 	336(2)	\\
0.16                   & 	16(1)		& 	334(2)	\\
0.21                   & 	11(1)		& 	332(2)	\\
0.31                   & 	17(1)		& 	327(2)	\\
0.42                   & 	13(1)		& 	327(1)	\\
0.52                   & 	12(1)		& 	325(1)	\\
0.67                   & 	15(1)		& 	316(2)	\\
0.81                   & 	13(1)		& 	319(2)	\\
1                      & 	8(1)		& 	309(2)	\\
\end{tabular}
\end{ruledtabular}
\end{table}

%\clearpage

\section{\label{Neuts} Neutron Diffraction}

\begin{figure}
\includegraphics[width=2.5 in]{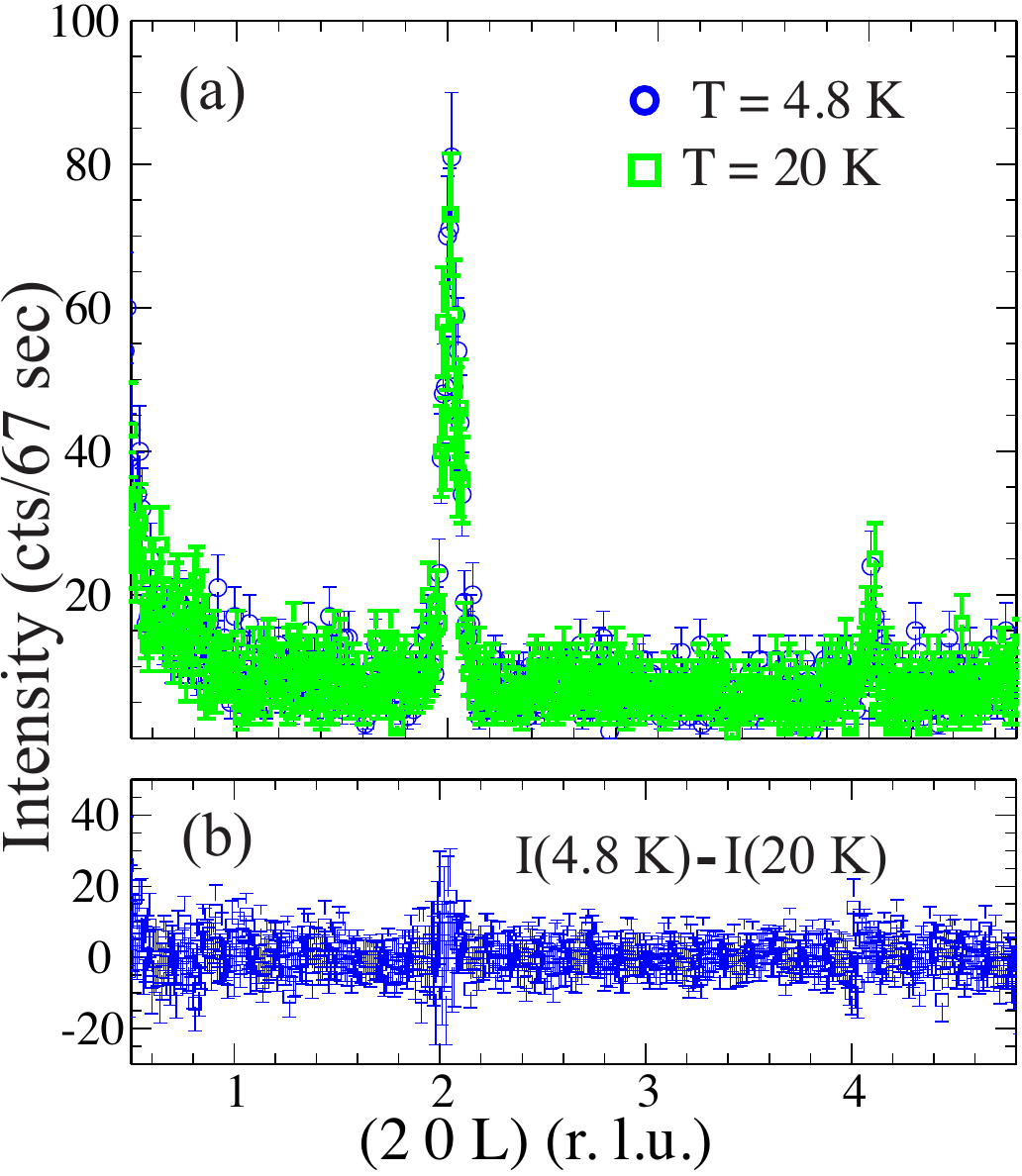}
\caption{  (a) Neutron-diffraction scan along $(2 0 L)$ reciprocal-lattice units (r.l.u.) from a single crystal with $x=0.16$ at $T=4.8$~K and at 20~K  showing the nuclear (202) and (204) reflections with no indication of new emerging peaks at the (201) or (203) positions which would be indicative of the A-type AFM ordering observed for the parent compound CaCo$_{1.86}$As$_2$. (b)~The difference in intensity between the scans at $T=4.8$ K and 20~K.}
\label{Fig:20L}
\end{figure}

Guided by the magnetic structure of the A-type antiferromagnetism exhibited by the parent compound CaCo$_{1.86}$As$_2$, we conducted a thorough search for new emerging magnetic Bragg reflections along the $(20L)$~r.l.u.\ (reciprocal-lattice unit)  direction~\cite{Quirinale2013}. In particular, for A-type antiferromagnetism it is expected to observe (201) and (203) magnetic reflections.  Figure~\ref{Fig:20L}(a) shows  scans along $(20L)$ for an $x=0.16$ crystal at base temperature and at 20 K, i.e.,  above the anomaly found in $\chi_{ab}$ [Fig.~\ref{Fig_M-T_all_separate}(a)].  As shown by the difference between these two scans in Fig.~\ref{Fig:20L}(b), no obvious peaks are observed over the whole $L$ range thus excluding A-type ordering.  In fact, much longer time counting longitudinal and transverse scans at the (201) position exclude an A-type phase at $T=4.8$ K assuming that the average ordered moment is not larger than our resolution of $\sim 0.05\,\mu_{\rm B}$ per transition-metal atom.

Additional scans along $(00L)$, $(10L)$, and $(\frac{1}{2}0L)$ at \mbox{$T=4.8$} and  20~K (not shown) do not show evidence of new emerging Bragg reflections that may indicate magnetic ordering that is different from A-type AFM ordering.  For instance, the absence of magnetic reflections along $(00L)$ eliminates the helical structure recently discovered in the related Sr(Co$_{1-x}$Ni$_x$)$_{2-y}$As$_2$ system in which adjacent FM layers (moments aligned in the plane) are rotated at a finite angle as they stack along the $c$~axis~\cite{Wilde2019}.  We note that examination of nuclear Bragg reflections do not show increased intensity as the temperature is lowered from 20~K to base temperature thus excluding FM ordering of the whole system assuming that the ordered moment is not larger than $\sim0.1\,\mu_{\rm B}$.

Similar experiments for the $x=0.11$ crystal with scans along $(20L),\, (00L),\, (H00),$ and $(10L)$  at base temperature and at 20 K were conducted (not shown) and also did not yield evidence of any of the various scenarios that were excluded above within our resolution for the $x=0.16$ crystal. 

\section{\label{Theory} Theory}

\subsection{Band-filling dependence of magnetic phases and their competition}

\begin{figure}[th]
	%\centering
	\includegraphics[width=2.in]{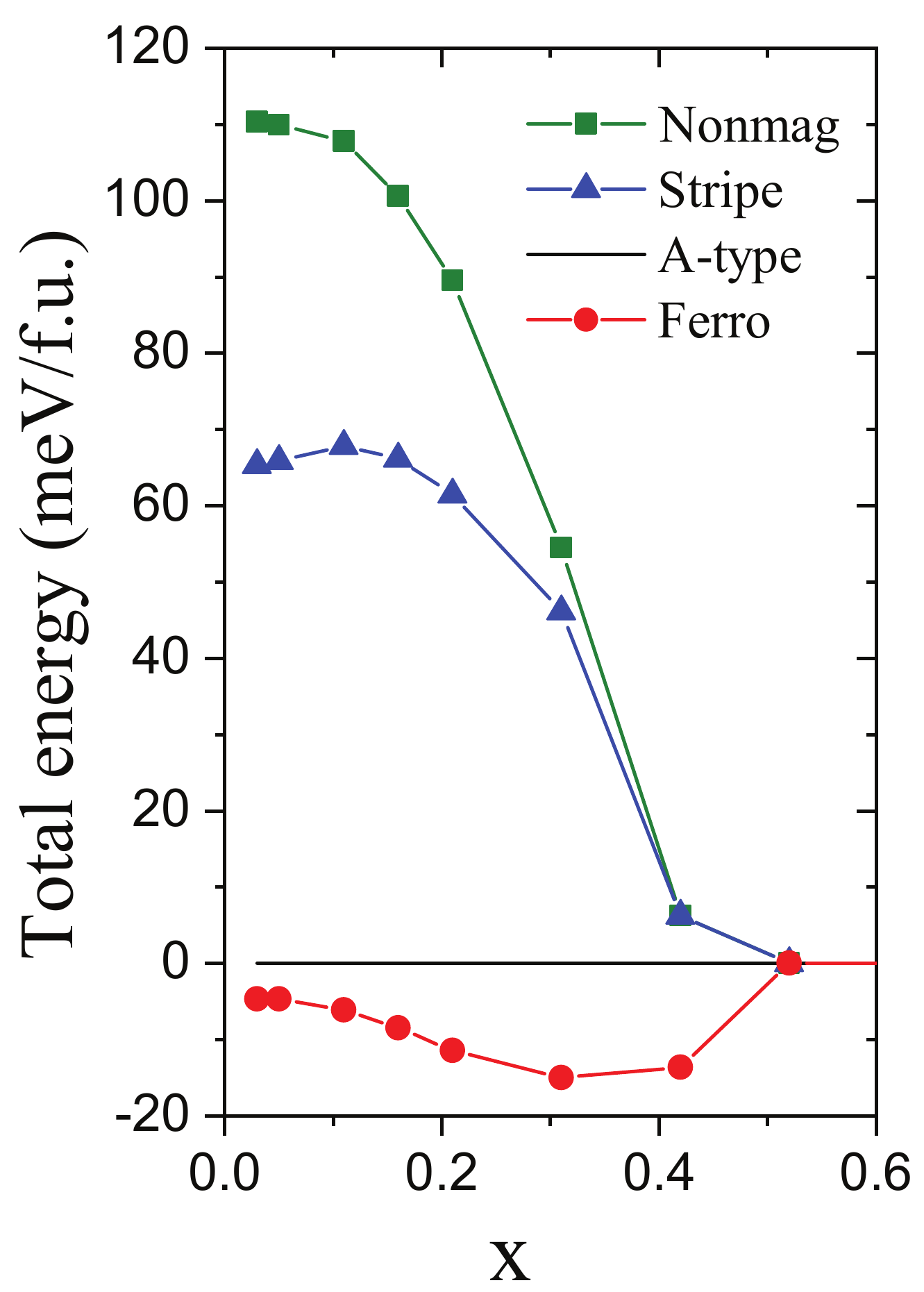}
	\caption{The effects of electron doping $x$ on the
                  magnetic energies of various magnetic states in the cT-phase CaCo$_{2}$As$_{2}$.
The energy of the A-type AFM phase is chosen as the reference zero.  The energies of nonmagnetic, stripe-AFM, and FM states are plotted.  The results are calculated using the experimental lattice parameters listed in Table~\ref{CrystalData}.}
	\label{fig:emag_vs_ne}
\end{figure}

Figure~\ref{fig:emag_vs_ne} shows the energies of four magnetic configurations, including ferromagnetic, A-type, stripe-AFM, and nonmagnetic states in Ca(Co$_{1-x}$Ni$_{x}$)$_2$As$_2$.
For stoichiometric CaCo$_{2}$As$_{2}$, FM and A-type AFM configurations have much lower energy than the stripe-AFM ordering, indicating FM intralayer moment alignment is favored over the stripe-AFM intralayer alignment, as observed.
The energy difference between the stripe-AFM phase and FM phase is much larger in cT-phase CaCo$_2$As$_2$ than in T-phase SrCo$_2$As$_2$~\cite{jayasekara2013}, mainly due to the different $c/a$-ratios in the T and cT phases.
As a result, stripe ordering or fluctuations are unlikely to occur in CaCo$_2$As$_2$-related compounds.

The energy difference between the FM and A-type AFM configurations is much smaller, suggesting that interlayer coupling is much weaker than intralayer coupling.
Interestingly, the stoichiometric CaCo$_{2}$As$_{2}$ FM state with $x=0$ even has a lower energy than the A-type AFM state found in CaCo$_{1.86}$As$_{2}$ from experiments.
This suggests that the intrinsic vacancies and the induced local structural relaxation present in experimental samples may play a
significant role in stabilizing the AFM interlayer coupling.

Compared to SrCo$_2$As$_2$, \cca\ shows a more stable magnetic ordering and the FM fluctuations exist in a much broader doping range in the Ni-doped compounds.  This is likely due to the lack of intralayer FM-AFM competition and the easy-axis anisotropy in CaCo$_2$As$_2$.
The much higher stripe-phase energy excludes stripe ordering or fluctuations in CaCo$_2$As$_2$, while both interlayer and intralayer FM-AFM competition coexist in SrCo$_2$As$_2$.
Moreover, unlike SrCo$_2$As$_2$, CaCo$_2$As$_2$ has an easy-axis magnetocrystalline anisotropy, which is essential to stabilize the long-range magnetic ordering in 2D systems or layered bulk systems with very weak interlayer couplings, according to the Mermin-Wagner theorem~\cite{Mermin1966}.

The calculated magnetocrystalline anisotropy energy in stoichiometric CaCo$_2$As$_2$ is $\Delta E =41~\mu$eV/Co with the easy direction along the $c$~axis, consistent with the easy $c$-axis anisotropy found in experiments on ${\rm CaCo_{1.86}As_2}$~\cite{Anand2014}.
For comparison, using the experimental value of the \mbox{$c$-axis} spin-flop field $H_{\rm SF} = 3.5$~T and the ordered moment $\mu=0.28~\mu_{\rm B}$/Co~\cite{Anand2014}, we obtain $\Delta E = \frac{1}{2}\mu H_{\rm SF} = 28~\mu$eV/Co.  The difference between the two values may be associated with the Co vacancies in ${\rm CaCo_{1.86}As_2}$ which were not accounted for in the theoretical calculation.  In addition, magnetic-dipole interactions~\cite{Johnston2016} between the Co spins in the simple-tetragonal Co structure  give a small XY anisotropy $\Delta E \approx -1.5\,\mu$eV/Co assuming Co point-dipole local moments.

\subsection{Density of States}

\begin{figure}[t]
	\includegraphics[width=2.in]{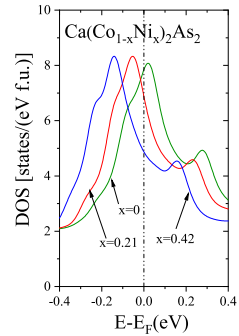}
	\caption{The effects of electron doping on the
                  density of states (DOS) versus energy $E$ relative to the Fermi energy $E_{\rm F}$ of Ca(Co$_{1-x}$Ni$_x)_2$As$_2$  for compositions $x=0$, 0.21, and 0.42. The DOS is expressed per formula unit, taking into account the twofold Zeeman degeneracy of the electron.  The DOS($E_{\rm F})$ is seen to peak near zero doping and to decrease upon electron doping.}
	\label{Fig:DOS}
\end{figure}

The influence of electron doping on the nonmagnetic density of electron states (DOS) is plotted versus energy~$E$ relative to the Fermi energy~$E_{\rm F}$ in Fig.~\ref{Fig:DOS}. The DOS in undoped CaCo$_2$As$_2$ peaks near $E_{\rm F}$ as previously reported~\cite{ueland2021arxiv}.  From Fig.~\ref{Fig:DOS}, a large amount of electron doping decreases DOS($E_\text{F}$) and suppresses the on-site magnetic moment according to a Stoner-like mechanism within a rigid-band picture, as also found previously for hole doping~\cite{ueland2021arxiv}.  Indeed, with electron doping, as shown in Fig.~\ref{fig:emag_vs_ne} the on-site magnetic moment of $3d$ atoms in Ca(Co$_{1-x}$Ni$_{x}$)$_2$As$_2$ vanishes at the Stoner transition composition $x\approx0.5$, which agrees well with the experimental fact that no spin fluctuations and magnetic ordering were observed with $x > 0.52$.
Interestingly, calculations also found that a smaller amount of electron doping promotes the FM interlayer coupling against the AFM one, consistent with the above experiments showing that electron doping suppresses the AFM interlayer coupling and induces strong FM fluctuations.

\begin{table}
\caption{\label{Tab:DOS(EF)} Calculated total density of states at the Fermi energy ${\cal D}(E_{\rm F})$ of Ca(Co$_{1-x}$Ni$_x)_2$As$_2$ per formula unit for both spin directions  for compositions $x=0$, 0.21, and 0.42.  Also listed are two similar electron-doping values~$q$ and corresponding ${\cal D}(E_{\rm F})$ inferred for \ccna\ from Table~\ref{Tab.heatcapacityfit}.}
\begin{ruledtabular}
\begin{tabular}{ccccc}	
$x$		& $q_{\rm calc}$	& ${\cal D}(E_{\rm F})_{\rm calc}$ 	&  $q_{\rm obs}$ 	& ${\cal D}(E_{\rm F})_{\rm obs}$ \\
		& (e/f.u.)			& (states/eV\,f.u.)				& (e/f.u.)			&  (states/eV\,f.u.)	\\
\hline
0		&  0				&	7.846						&  0.06			& 13.9(2) 			\\
0.21		& 0.42 			&	6.754						&  0.50			& 12.55(2)	 				   \\
0.42		& 0.84			&	4.809						&  1.15			&  7.95(3)	   \\
    \end{tabular}
\end{ruledtabular}
\end{table}

The values of the calculated densities of states at the Fermi energy ${\cal D}(E_{\rm F})$ from Fig.~\ref{Fig:DOS} are listed for $x=0$, 0.21, and 0.42 in Table~\ref{Tab:DOS(EF)}.  Three measured values for approximately the same electron-doping levels from Table~\ref{Tab.heatcapacityfit} are listed for comparison.  The observed values are roughly a factor of two larger than the bare calculated values, suggesting significant enhancements from electron-phonon and/or electron-electron interactions.

\section{\label{ConcRem} Summary}

\begin{figure}[t]
\includegraphics[width = 3in]{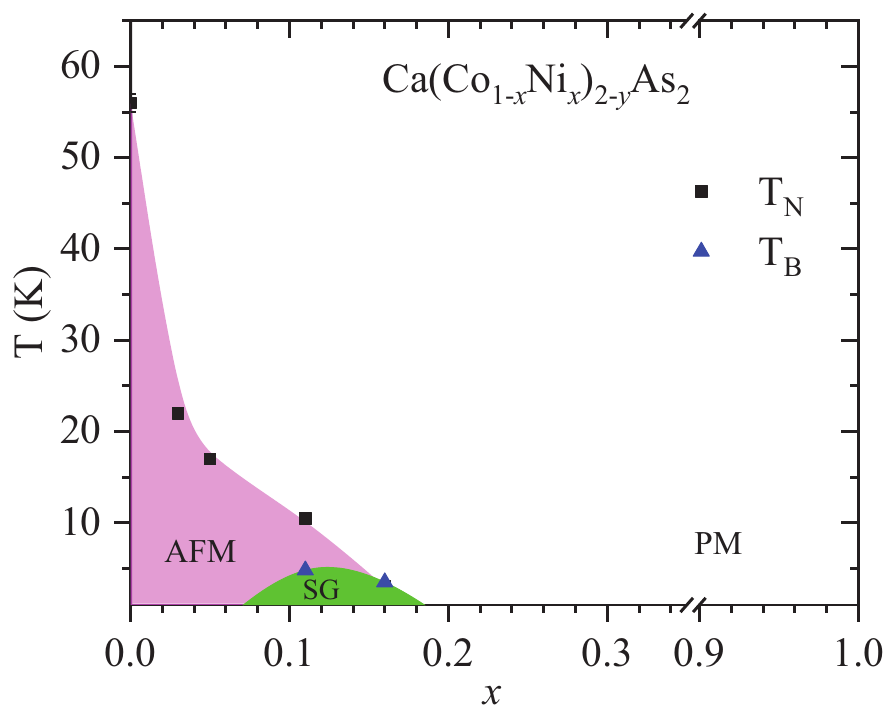}
\caption{Magnetic phase diagram in the $T$-$x$ plane  of the \ccna\ system obtained from $\chi(T)$ and $M(H)$-isotherm measurements. Here AFM, SG, and PM stand for antiferromagnetic, spin-glass, and paramagnetic states, respectively.  The suppression of AFM order by the development of strong FM fluctuations in \ccna\ crystals suggest the presence of a nearby FM QCP at $x \approx 0.2$. Note the  break in the abscissa scale between $x = 0.34$ and $x= 0.9$.}
\label{Fig_phase_diagram}
\end{figure}

The magnetic phase diagram in the $T$-$x$ plane of the \ccna\ system based on the $\chi(T)$, $M(H)$ isotherm and $M(t)$ isotherm and isofield measurements is shown in Fig.~\ref{Fig_phase_diagram}. The A-type AFM ordering observed at $T_{\rm N} = 52$~K in the highly-frustrated itinerant parent compound \cca\ is gradually suppressed with Ni substitution and vanishes for $x > 0.11$. Ni substitution strongly enhances the FM spin fluctuations in this system and as a result $T_{\rm N}$ rapidly drops to 22~K with only a 3\% Ni~substitution along with a strong enhancement of the  Sommerfeld electronic heat-capacity coefficient.

The FM fluctuations increase further in the $x = 0.11$ and 0.16 crystals and compete with the AFM ordering in these systems until the formation of a low-$T$ spin-glass state below a blocking temperature $T_{\rm B}\sim 5$~K\@.  With a further increase in Ni concentration quasi-one-dimensional $c$-axis FM spin fluctuations dominate in the $x = 0.21$ and 0.31 crystals where $\chi_c$ becomes much larger than $\chi_{ab}$ in the low-$T$ region.  Small static FM moments at $T=2$~K were observed for $x=0.11$, 0.16, 0.21, and 0.31 from magnetization versus field isotherms, with a maximum value of 0.12~$\mu_{\rm B}$/f.u.\ for $x=0.16$.

DFT calculations confirm that FM fluctuations are enhanced by Ni substitutions for Co in \cca.  The FM spin fluctuations are observed over a wide composition range from $x = 0.11$ to $x = 0.52$ at which a Stoner transition to a nonmagnetic state occurs as previously observed in the hole-doped Ca(Co$_{1-x}$Fe$_x)_{2-y}$As$_2$ system~\cite{ueland2021arxiv}. 

The results suggest the presence of a FM quantum-critical point at $x \approx 0.20$, which is unusual in such materials. However, a FM QCP can be avoided if preempted by an AFM transition~\cite{Brando2008, Lengyel2015, Hamann2019}. If FM critical fluctuations persist to low temperatures, a novel ground state such as $p$-wave superconductivity is conceivable~\cite{Fay1980, Kuzmin2000, Mochizuki2005, Hattori2012} with an onset below our low-$T$ measurement limit of 1.8~K\@.

%\clearpage

\acknowledgments

This research was supported by the U.S. Department of Energy, Office of Basic Energy Sciences, Division of Materials Sciences and Engineering.  Ames Laboratory is operated for the U.S. Department of Energy by Iowa State University under Contract No.~DE-AC02-07CH11358.

%\clearpage

\end{document}